\documentclass[fleqn,usenatbib]{mnras}

\usepackage{newtxtext,newtxmath}

\usepackage[T1]{fontenc}
\usepackage{ae,aecompl}
\usepackage{todonotes}

\usepackage{graphicx}	
\usepackage{amsmath}	
\usepackage{amssymb}	
\usepackage{hyperref}

\newcommand{\HII}{\rm H~{\sc ii }}

\title[Neutral Island statistics]{Neutral island statistics during reionization from 21-cm tomography}

\author[S. K. Giri et al.]{
Sambit K. Giri$^{1}$\thanks{E-mail: \href{mailto:sambit.giri@astro.su.se}{sambit.giri@astro.su.se}},
Garrelt Mellema$^{1}$,
Thomas Aldheimer$^{1}$,
Keri L. Dixon$^{2,3}$
\newauthor
~and Ilian T. Iliev$^{3}$
\\
$^{1}$
Department of Astronomy and Oskar Klein Center, Stockholm University, AlbaNova, SE-106 91 Stockholm, Sweden\\
$^{2}$
New York University Abu Dhabi, PO Box 129188, Saadiyat Island, Abu Dhabi, UAE\\
$^{3}$
Astronomy Centre, Department of Physics \& Astronomy, Pevensey II Building, University of Sussex, Falmer, Brighton BN1 9QH, UK
}

\date{Accepted 2019 August 8. Received 2019 July 10; in original form 2019 February 26}

\pubyear{2019}

\begin{document}
\label{firstpage}
\pagerange{\pageref{firstpage}--\pageref{lastpage}}
\maketitle

\begin{abstract}
We present the prospects of extracting information about the Epoch of Reionization 
by identifying the remaining neutral regions, referred to as islands, in tomographic observations of the redshifted 21-cm signal. Using simulated data sets we show that at late times the 21-cm power spectrum is fairly insensitive to the details of the reionization process but that the properties of the neutral islands can distinguish between different reionization scenarios. We compare the properties of these islands with those of ionized bubbles. At equivalent volume filling fractions, neutral islands tend to be fewer in number but larger compared to the ionized bubbles. In addition, the evolution of the size distribution of neutral islands is found to be slower than that of the ionized bubbles and also their percolation behaviour differs substantially. Even though the neutral islands are relatively rare, they will be easier to identify in observations with the low frequency component of the Square Kilometre Array (SKA-Low) due to their larger size and the lower noise levels at lower redshifts. The size distribution of neutral islands at the late stages of reionization is found to depend on the source properties, such as the ionizing efficiency of the sources and their minimum mass. We find the longest line of sight through a neutral region to be more than 100 comoving Mpc until very late stages (90-95 per cent reionized), which may have relevance for the long absorption trough at $z=5.6-5.8$ in the spectrum of quasar ULAS J0148+0600.
\end{abstract}



\section{Introduction}
\label{sec:intro}
During the Epoch of Reionization (EoR), ionizing photons emitted by the first generations of galaxies (re)ionized the neutral hydrogen in the intergalactic medium \citep[IGM, see][for a recent review]{Dayal2018EarlyEffects}. The spin-flip transition of the neutral hydrogen produces a signal with an intrinsic wavelength of 21-cm \citep[e.g.,][]{furlanetto200421,Mellema2013ReionizationArray} and imaging this signal at different redshifts has the potential to map the distribution of neutral hydrogen. By following the evolution of the signal with redshift we can extract information about the reionization 
process and the properties of the sources that caused it \citep[e.g.,][]{Furlanetto2004TheReionization,morales2010reionization,2012MNRAS.423.2222I}.

Indirect observations of the state of the IGM such as the Lyman-$\alpha$ absorption in the spectra of high-redshift quasars and the Thomson scattering optical depth of the cosmic microwave background (CMB) radiation have constrained majority of reionization to have happened in the redshift range $z \approx 6-10$ 
\citep[e.g.,][]{Fan2006ObservationalReionization, Mitra2015CosmicPlanck,2016A&A...594A..13P}. The recent claimed detection of the global 21-cm signal from $z\approx 17$ by the EDGES\footnote{Experiment to Detect the Global EoR Signature} team illustrates the potential of the signal to reveal unique information about these early times \citep{Bowman2018AnSpectrum}. A range of interferometric radio telescopes, such as the Giant Metrewave Radio Telescope \citep[GMRT; e.g.,][]{Paciga2011The8.6}, the Low-Frequency Array \citep[LOFAR; e.g.,][]{Harker2010PowerCase}, the Murchison Widefield Array \citep[MWA; e.g.,][]{Lonsdale2009TheOverview} and the Precision Array for Probing the Epoch of Reionization \citep[PAPER; e.g.,][]{Parsons2010TheResults} are attempting to detect the power spectrum of the 21-cm signal and observations have resulted in upper limits \citep{Jacobs2015MultiredshiftPAPER, Patil2017UpperLOFAR} in spite of major challenges due to the strong foreground signals.

Future radio telescopes, such as Hydrogen Epoch of Reionization Array \citep[HERA;][]{deboer2017hydrogen} and the Square Kilometre Array \citep[SKA;][]{Mellema2013ReionizationArray}, will have even greater sensitivity to observe the 21-cm signal from high $z$. The low frequency component of the SKA (SKA-Low) plans to go beyond a statistical detection of EoR by producing images. As the 21-cm signal is a spectral line, an image produced at a certain frequency will map the Universe at a particular redshift. A sequence of such images varying with frequency is known as the 21-cm tomography \citep[e.g.,][]{furlanetto200421} and will map the evolution of the distribution of neutral hydrogen in the Universe during the EoR.


In \citet{Giri2018BubbleTomography}, we showed how extracting the size distribution of ionized regions from the tomographic data sets can be used to study the EoR. These ionized regions are often referred to as `bubbles'. The ionized bubbles have direct physical correlation with the reionization process as the evolution of their sizes and morphology is driven by the properties of the sources located inside them \citep[e.g.,][]{Furlanetto2004TheReionization,McQuinn2007TheReionization,Friedrich2011TopologyReionization}. However, this connection becomes less clear when a large fraction of the Universe has reionized and photons from any source start to affect regions at large distances due to the negligible optical depth of the ionized gas. 

During the end stages of reionization, the remaining neutral regions, which we will refer to as `islands', will appear as isolated structures in the 21-cm images. These neutral islands are expected to be relatively large \citep[e.g.,][]{Kulkarni2019Large5.5,Keating2019LongHydrogen} and at the lower $z$ the signal-to-noise ratio (SNR) of the images will be better. Therefore these neutral islands 
should be relatively easy to identify in the 21-cm tomographic images. This will be especially true during the first years of operation of SKA-Low when only data sets of limited integration time and thus with higher noise levels will be available. In those it may be easier to detect neutral islands at low redshifts than ionized bubbles at high redshifts. In this paper, we study the information contained in the sizes and morphology of these neutral islands. 


Late neutral islands have not received as much attention as the ionized bubbles in previous work on the 21-cm signal from the EoR. Occasionally, size distributions of neutral regions were presented \citep[e.g.][]{2007ApJ...669..663M, Iliev2014SimulatingEnough} but always within studies focusing on the ionized bubbles. An exception to this was \citet{Zaroubi2012ImagingLOFAR} who pointed out the presence of large neutral regions during the later stages of reionization and also made the point that these may be easier to image than early ionized regions. However, these authors did not explore the statistical properties of the neutral islands. An explanation for the lack of studies is that the late stages of reionization are more difficult to model accurately, especially for semi-numerical codes relying on the excursion set formalism such as \textsc{\small 21cmFAST} \citep{Mesinger201121cmfast:Signal}. Once ionized bubbles substantially overlap these codes suffer from a lack of photon conservation \citep[e.g.,][]{Choudhury2018PhotonReionization}. This drawback motivated \citet{Xu2013AnReionization} to develop a modified formalism to improve the calculations of the late stages. The \textsc{\small islandFAST} simulation code uses that formalism to simulate the end of reionization and was used to study statistical properties of neutral islands \citep{Xu2017IslandFAST:Reionization}.


Here we use fully numerical, radiative transfer, reionization simulations to investigate the properties of neutral islands in both the simulation and observational frameworks. In the simulation framework, we strive to understand the properties of the EoR that affect the summary statistics such as the size distribution and the Euler characteristic ($\chi$). In the observational framework, we mimic the telescope effects, such as the system noise and the limited resolution, and add them to the simulated 21-cm signal. These effects will affect our ability to extract EoR information from the observed summary statistics.

In the next section, we explain how we construct the 21-cm signal from the simulation data. We present our identification and analysis methods of the neutral islands in Section~\ref{sec:method}. This section is followed by a section describing and discussing our results. Section~\ref{sec:conclusions} provides a summary and our conclusions.

\section{Redshifted 21-cm signal}

\subsection{Simulating reionization}
\label{sec:sims}
As no 21-cm images from the EoR have yet been observed, we will use mock observations for our study. These are constructed to mimic the 21-cm observations by the future SKA-Low interferometer.
We calculate the cosmological 21-cm signal from the results of a fully numerical reionization simulation. These simulations consist of two stages. First, the evolution of the matter distribution is calculated using the $N$-body code \textsc{\small CUBEP$^3$M} \citep{2013MNRAS.436..540H}. Next, the ionization structure of the matter is calculated in post-processing mode with \textsc{\small C$^2$-RAY} \citep{2006NewA...11..374M}, a radiative transfer simulation code. Our simulation methodology has been described in more detail in previous papers \citep[e.g.,][]{2006MNRAS.372..679M, 2012MNRAS.424.1877D, Iliev2014SimulatingEnough}.
 The cosmological parameters used here are $\Omega_\mathrm{m}$=0.27, $\Omega_\mathrm{k}$=0, $\Omega_\mathrm{b}$=0.044, $h=0.7$, $n_\mathrm{s}=0.96$ and $\sigma_8$=0.8. These values are consistent with the results from \textit{Wilkinson Microwave Anisotropy Probe} (WMAP) \citep{2011ApJS..192...18K} and Planck \citep{2016A&A...594A..13P,PlanckCollaboration2018PlanckParameters}.
 
 \begin{figure*}
  \centering
  \includegraphics[width=1.0\textwidth]{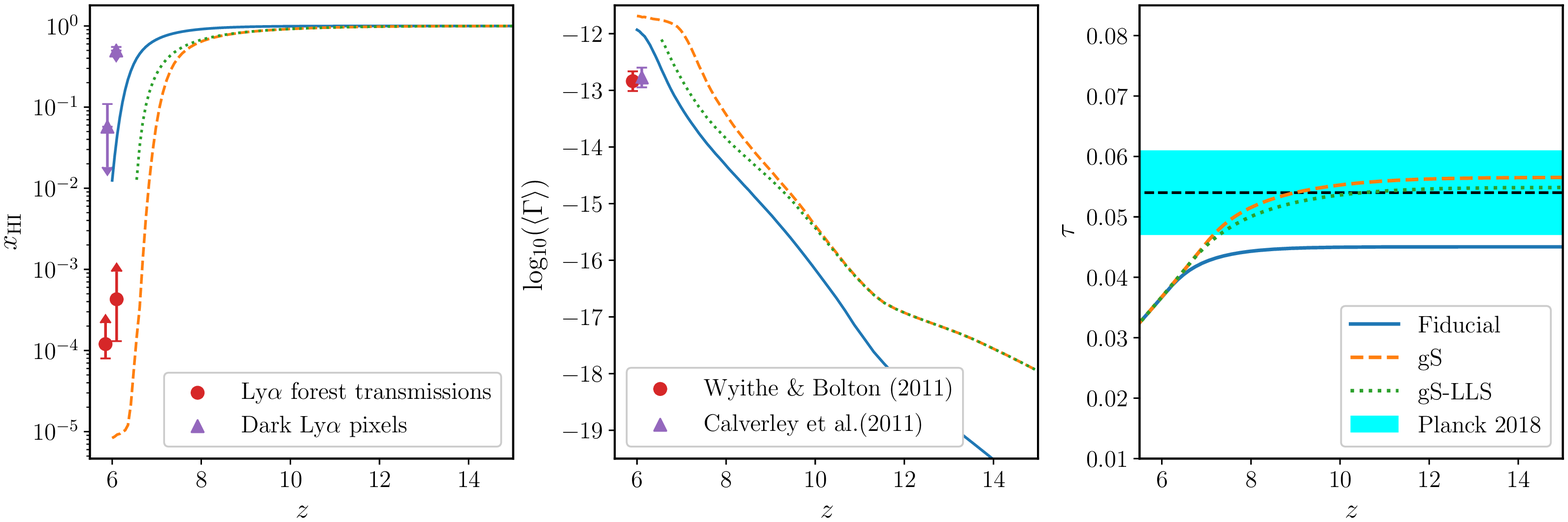}
  \caption{Our reionization simulations with different source models compared to the observational constraints. \textit{Left panel:} The volume weighted average neutral fraction ($x_\mathrm{HI}$) is compared to the observational constraints with Ly$\alpha$ forest transmissions \citep[red circles;][]{Fan2006ConstrainingQuasars} and dark Ly$\alpha$ pixels \citep[violet triangles;][]{McGreer2011TheAt,McGreer2015Model-independentZ6}. \textit{Middle panel:} The mean volume-weighted hydrogen photoionization rate ($\langle \Gamma \rangle$) compared to the observational constraint of \citet[][red circle]{Wyithe2011Near-zoneQuasars} and \citet[][violet triangle]{Calverley2011MeasurementsEffect}. \textit{Right panel:} We plot the Thomson scattering optical depth ($\tau$) calculated from our simulations. The constraints on $\tau$ put by the \textit{Planck 2018} data \citep{PlanckCollaboration2018PlanckParameters} is shown with black dashed line and the shaded region represents the 1$\sigma$ interval.}
  \label{fig:history_source_models}
\end{figure*}

The sources of ionizing radiation are associated with dark matter haloes that have been identified in the $N$-body results using the spherical overdensity halo-finder discussed in \citet{2013MNRAS.433.1230W}. One of the few papers presenting a size distribution of neutral islands, \citet{Iliev2014SimulatingEnough}, showed that the larger neutral islands are rare and therefore require large simulation volumes to sample them properly. Therefore we use a cosmological volume of 714 comoving Mpc (cMpc) containing $6912^3$ particles.
{We used a higher resolution version of the same simulation volume to study the squeezed-limit bispectrum of the 21-cm signal in \citet{Giri2019Position-dependentReionization}.}
This choice for the $N$-body simulation allows us to identify haloes with masses $M_\mathrm{halo} \geq 10^9 \mathrm{M}_\odot$. We call haloes in this mass range high-mass atomic-cooling haloes (HMACHs). Gas in these haloes is able to cool through excitation of atomic hydrogen and they are relatively unaffected by radiative feedback \citep{2018MNRAS.473...38S}. Haloes in the mass range 10$^8 \mathrm{M}_\odot$ < $M_\mathrm{halo}$ < 10$^9 \mathrm{M}_\odot$ are also able to cool atomically, but are affected by radiative feedback. These are called low-mass atomic-cooling haloes (LMACHs). As these are not resolved in our $N$-body simulation, we add them using a sub-grid model as described in \citet{2015MNRAS.450.1486A}. Haloes with masses below $10^8 \mathrm{M}_\odot$ can only cool through molecular hydrogen and are not considered here.

\begin{table*}
	\centering
	\caption{Reionization simulation parameters of various source models.}
	\label{tab:sim_list}
	\begin{tabular}{lcccccccc} 
		\hline
		Label  & Box size (Mpc) & Mesh & $g_{\gamma}$ (HMACHS)  & $g_{\gamma}$ (LMACHS) & LLS \\
		\hline
		Fiducial & 714 & 300$^3$ & 1.7 & 0   & $R_\mathrm{mfp}=57$~cMpc \\
		gS       & 714 & 300$^3$ & 1.7 & equation~(\ref{eq:gs}) & $R_\mathrm{mfp}=57$~cMpc\\
		gS-LLS   & 714 & 300$^3$ & 1.7 & equation~(\ref{eq:gs}) & Position and redshift dependent\\
		\hline
	\end{tabular}
\end{table*}

As they are affected differently by radiative feedback, the LMACHs and HMACHs are treated differently. In this study, we consider three different source models. An overview of different approaches to modelling sources is given in \citet{2016MNRAS.456.3011D}. The source efficiencies are characterized by the quantity $g_\gamma$ which gives the number of ionizing photons per baryon escaping from the sources into the IGM per 10$^7$ years. In all our simulations the HMACHs are assigned $g_\gamma$(HMACH)=1.7 but the LMACHs are treated differently. In the `Fiducial' simulation, the LMACHs are inactive throughout the reionization history. In the other two simulations (labelled `gS' and `gS-LLS') LMACHs have the same efficiency as HMACHs in neutral regions and suffer from a mass-dependent suppression in ionized regions, according to  \citep[][]{2016MNRAS.456.3011D},
\begin{equation}
g_\gamma (\mathrm{LMACH}) = 
\begin{cases} 
g_\gamma (\mathrm{HMACH}), \quad \ \text{if $x_\mathrm{HII}<x_\mathrm{HII,threshold}$}. \\
g_\gamma (\mathrm{HMACH}) \left[\frac{M_\mathrm{halo}}{9\times 10^8 \mathrm{M}_\odot} - \frac{1}{9}\right], \\
\qquad \qquad \qquad \ \ \ \text{if $x_\mathrm{HII}\geq x_\mathrm{HII,threshold}$}. 
\end{cases}
\label{eq:gs}
\end{equation}
with $x_\mathrm{HII,threshold}=0.1$. We call our simulations with the above mass-dependent suppression as gradual suppression (gS) models. The above prescription is motivated by the radiation-hydrodynamic simulations of \citet{2009ApJ...693..984W} and \citet{2018MNRAS.473...38S}. Comparing the results of our Fiducial case with those for the gS case will probe the impact of using different source populations.

Apart from the source efficiency, the size and shape of the late neutral regions will be affected by opacity of the ionized IGM, which determines how far ionizing photons can travel. This opacity is dominated by the presence of Lyman-limit systems (LLS), small and dense structures which are opaque to ionizing photons. In both the Fiducial and gS simulations, we include their effect by imposing a maximum distance of 57~cMpc that ionizing photons can travel. In the third simulation `gS-LLS', the effects of LLS is modelled by adding an extra opacity component to the IGM. This additional opacity is distributed over the cells to ensure that on average an optical depth of unity at the Lyman limit is reached after photons have travelled a distance $\ell_\mathrm{mfp}(z)$. However, the actual opacity in a cell depends on the sum of the cross sections of the haloes contained in that cell where the radius of a halo is taken to be its virial radius. This position-dependent approach is motivated by the fact that LLS are associated with the denser areas of the IGM. The value of $\ell_\mathrm{mfp}$ at a redshift $z$ is found from an extrapolation of the redshift dependence of this quantity for $z<6$ taken from \citet{2010ApJ...721.1448S}. See \citet{2016MNRAS.458..135S} for more details and comparisons of different approaches for LLS.
By comparing the results for gS and gS-LLS, we will be able to gauge how sensitive they are to the choice of implementing the opacity of the ionized IGM. We list the parameters of the reionization simulations used in Table~\ref{tab:sim_list}. 

In Fig.~\ref{fig:history_source_models}, we compare the three scenarios against observational constraints. In the right most panel, we see that all three models are close to the latest estimates for the electron scattering optical depth, although our Fiducial model actually falls slightly outside the 1-$\sigma$ regime. The average photoionization rate $\Gamma$ is shown in the middle panel. All three scenarios overshoot the measured values around $z=6$. This means either that our models underestimate the mean free path of ionizing photons or that our sources are too bright. Comparing the gS and the gS-LLS cases shows that our LLS model indeed brings down the value of $\Gamma$, however, not enough to reproduce the measured values. As the LLS model is based on the measured value of the mean free path around $z=6$, this suggests that our sources are too bright. 

The left-most panel of Fig.~\ref{fig:history_source_models} shows the evolution of the mean neutral fraction. Also these results indicate that our sources are too bright as reionization proceeds too quickly. The gS model is the only one which fully completes and then the remaining neutral fraction is lower than the one measured from the quasar spectra, consistent with the higher value of $\Gamma$.

Similar results were also shown in \citet{2016MNRAS.456.3011D} and we conclude that our simulations do not reproduce the proper transition to the Ly$\alpha$ forest measurements of the post-reionization Universe. In fact, very few simulations do and it seems that a rather careful choice for the evolution of the emissivity is required to obtain the correct behaviour \citep[see e.g.,][]{Kulkarni2019Large5.5,Keating2019LongHydrogen}. Here we will present the results for the three scenarios to show that the studying the properties of late neutral islands is useful and important. The differences between the results for the three models will be instructive to gauge how large the impact of differences in source properties or the treatment of the IGM opacity will be.

\subsection{Observed signal}
\label{sec:observed_signal}
\subsubsection{Brightness temperature}
The differential brightness temperature of the observed 21-cm signal is given as \citep[e.g.,][]{Mellema2013ReionizationArray}
\begin{eqnarray}
\delta T_\mathrm{b} \approx 27 x_\mathrm{HI} (1 + \delta)\left( \frac{1+z}{10} \right)^\frac{1}{2}
\left( 1 -\frac{T_\mathrm{CMB}(z)}{T_\mathrm{s}} \right)\nonumber\\
\left(\frac{\Omega_\mathrm{b}}{0.044}\frac{h}{0.7}\right)
\left(\frac{\Omega_\mathrm{m}}{0.27} \right)^{-\frac{1}{2}} 
\left(\frac{1-Y_\mathrm{p}}{1-0.248}\right)
\mathrm{mK}\ ,
\label{eq:dTb}
\end{eqnarray}
where $x_\mathrm{HI}$ and $\delta$ are the fraction of neutral hydrogen and the density fluctuation respectively. $T_\mathrm{s}$ is the excitation temperature of the two spin states of the neutral hydrogen, known as the spin temperature. $T_\mathrm{CMB}(z)$ is the CMB temperature at redshift $z$ and $Y_\mathrm{p}$ is the primordial helium abundance. 

From equation~(\ref{eq:dTb}), it is clear that there will be no signal when $T_\mathrm{CMB} \approx T_\mathrm{s}$. It is expected that the spin temperature decoupled from $T_\mathrm{CMB}$ and approached the gas temperature due to the Wouthuysen-Field effect during the EoR \citep{1997ApJ...475..429M}. This makes the 21-cm signal observable. When the gas temperature is below $T_\mathrm{CMB}$,  the signal is seen in absorption and when it is above, it is seen in emission. In this study, we assume the high spin temperature limit, $T_\mathrm{s}\gg$ $T_\mathrm{CMB}$, in which case the signal becomes independent of the value of $T_\mathrm{s}$. 
The high spin temperature limit is generally considered to be a valid assumption for the later stages of reionization as even relatively low levels of X-ray radiation can raise the gas temperature above the CMB temperature \citep[e.g.,][]{2007MNRAS.376.1680P,2012RPPh...75h6901P}. In this case, regions with $\delta T_\mathrm{b}=0$ can be easily identified as ionized regions. Note however that interferometers do not yield the absolute signal due the absence of zero baselines and thus identification of ionized regions is less trivial in such data.  We discuss this further in Section~\ref{sec:telescope_effects} below.

Using equation~(\ref{eq:dTb}) we construct three-dimensional 21-cm data sets at a given redshift $z$ from the neutral fraction and density fields produced by \textsc{\small C$^2$-RAY} and \textsc{\small CUBEP$^3$M} respectively. This 3D data set represents an image cube where one of the axes is taken to be the frequency axis and the other two represent position in the sky. We disregard any evolution of the signal along the frequency direction, i.e.\ we use coeval cubes, not light cone cubes.
{We do this to minimise the effects of sample variance but note that real data will be light cone data in which evolution along the frequency direction will need to be considered. See \citet{Giri2018BubbleTomography} for a study of the impact of the light cone effect on bubble size statistics.}

The intrinsic resolution of our data set is $\Delta x=2.38$ cMpc that corresponds to angular scales
\begin{equation}
\Delta \theta = \frac{\Delta x}{D_\mathrm{c}(z)}\,,
\end{equation}
where $D_\mathrm{c}(z)$ is the comoving distance to redshift $z$, and frequency scales
\begin{equation}
\Delta \nu = \frac{\nu_0 H(z) \Delta x}{c(1+z)^2}\,,
\end{equation}
where $\nu_0$ is the rest frequency of the 21-cm line, $H(z)$ the Hubble parameter at redshift $z$ and $c$ the speed of light. For $z=7$, these equations give values $\Delta \theta=0.914$ arcmin and $\Delta \nu=0.141$~MHz.

\subsubsection{Adding telescope effects}
\label{sec:telescope_effects}
We follow the approach presented in \citet{Giri2018OptimalObservations} to add telescope effects to the simulated 21-cm signal. We model the effects using the current available configuration\footnote{The configuration is described in document SKA-TEL-SKO-0000557 Rev 1 and the positions of the individual stations in document SKA-TEL-SKO-0000422 Rev 2, both retrievable at \url{https://astronomers.skatelescope.org/documents/}} of the first phase of SKA-Low (hereafter SKA1-Low), which has 512 antennae stations each with a diameter of 35 m. 

\begin{table}
	\centering
	\caption{The parameters used in this study to model the telescope properties.}
	\label{tab:telescope_param}
	\begin{tabular}{lccccc} 
		\hline
		Parameters & Values \\
		\hline
        Observation time ($t_\mathrm{int}$)& 1000 h  \\
        System temperature ($T_\mathrm{sys}$) & $60 (\frac{\nu}{300\mathrm{MHz}})^{-2.55}$ K  \\
		Effective collecting area ($A_\mathrm{D}$) & 962 $\mathrm{m}^2$  \\
        Critical frequency ($\nu_\mathrm{c}$) & 110 MHz \\
        Declination & -30$^\circ$ \\
        Observation hour per day & 6 hours \\
        Signal integration time & 10 seconds \\
		\hline
	\end{tabular}
\end{table}

\paragraph{Limited resolution} A radio interferometric telescope produces a sky signal in so-called $uv$ space, which is the Fourier transform of the real sky image, see \citet{rohlfs2013tools} for details about observation with radio telescopes. The shortest baseline of the radio antennae corresponds to the largest scale present in the image while the largest baseline corresponds to the smallest scale, normally referred to as the resolution of the image. The SKA1-Low configuration includes baselines long enough to match the resolution of our simulation. However, it will be sparsely filled beyond a diameter of 2 km around the centre and therefore such high resolution observations will be very noisy. We therefore only use baselines shorter than 2~km and we approximate the synthesized beam by a Gaussian with a full-width half maximum (FWHM) given by equation~(4) of \cite{Giri2018BubbleTomography}. To implement the effect of the shortest baseline we subtract its average value from each image. In our simulated 21-cm signal data set, one axis is the frequency (or redshift) axis. To achieve isotropic resolution in our data set we convolve the signal along the frequency axis with a tophat filter of the same width as the Gaussian beam.

\paragraph{System noise}
In order to determine the noise map, we start with the value of its standard deviation, $\sigma$, calculated using equation~(4) from \citet{Giri2018OptimalObservations}. The values defining the observational setup are given in Table~\ref{tab:telescope_param}. The system noise is Gaussian in the $uv$ space. We therefore fill two arrays (amplitude and phase) of the same size as the signal data set with Gaussian random numbers with standard deviation $\sigma$.  We create the $uv$ tracks from our observational setup using the technique in \citet{Ghara2017ImagingSKA} and \citet{Giri2018OptimalObservations}. The pixels in the noise amplitude and phase arrays are set to zero if there are no $uv$ tracks passing through them or otherwise divided by $\sqrt{n_{uv}}$, where $\sqrt{n_{uv}}$ is the number of $uv$ tracks passing through that pixel. After this step, the arrays are inverse Fourier transformed to obtain a noise map in real space. This noise map is then added to the signal map.


\section{Methods}
\label{sec:method}
\subsection{Structure identification}
\label{sec:method_ident}
Ionized regions do not produce a 21-cm signal and will therefore appear as holes (regions with no signal) in the 21-cm signal maps. If we can identify the ionized regions in the tomographic observations, then we obtain the neutral islands as the complementary field. However this identification is not trivial. Due to the absence of the zero baseline in interferometric observations, we are incapable of recording the absolute value of the signal. The presence of noise and residual foregrounds will further complicate the identification process. For this study we will assume that the foreground mitigation procedure is perfect and will only consider the impact of noise, finite resolution and the inability to measure the absolute signal.

In \citet{Giri2018OptimalObservations}, we presented a method to identify the ionized region in noisy 21-cm observation during the EoR, called superpixels. Below we briefly describe the procedure to create the superpixels and the criteria used to identify superpixels containing ionized regions. The neutral regions will be the complimentary field of the ionized regions.

\subsubsection{Superpixels}
The morphology of the structures in the 21-cm observations is expected to be complex. The ionized regions will grow and overlap and form the so called percolation cluster \citep[][]{Iliev2006SimulatingReionization,Furlanetto2016ReionizationTheory}. Therefore identifying the entire structure will be difficult. However locally the structures will be less complex and can be identified. The superpixels method groups image-pixels with similar properties into superpixels. There exist several methods to create the superpixels \citep[e.g.,][]{Wei2016SuperpixelHierarchy}. We will use the method introduced by \citet{Achanta2012SLICMethods}, which generates a centroidal Voronoi tessellation of a given data set. The distance metric used for the Voronoi tessellation is a modified form of the Euclidean distance that includes a contribution from the pixel intensities so that pixels with similar intensities will be found to be closer to each other \citep[equation~(6) in ][]{Giri2018OptimalObservations}. For a detailed description of the procedure, see \citet{Giri2018OptimalObservations}.

\subsubsection{Stitching criteria}
Once we have grouped the pixels together into superpixels, we need to classify these into ionized and neutral ones. For this we use the probability distribution function (PDF) of the superpixels. Since several pixels contribute to a superpixel, the signal to noise for a superpixel is higher than for single pixels. As we showed in \citet{Giri2018OptimalObservations}, we can recover the intrinsic PDF of the 21-cm signal this way, see fig.~7 of that paper. During most of reionization, the 21-cm signal PDF is bimodal, which makes it easy to identify neutral and ionized regions. However, during the early stages, the observed PDF will be unimodal, although still asymmetric, which can also be inferred from the the non-zero value of the skewness of the signal \citep[e.g.,][]{Iliev2006SimulatingReionization,Watkinson2015TheMoments}. In this case, we use the maximum deviation method \citep{Rosin2001UnimodalThresholding} to use the asymmetry of the PDF to classify the superpixels as ionized. The final output after the stitching is a binary field where the pixels of our region of interest is unity and zero elsewhere.

\subsection{Structure analysis}
\label{sec:structure_analysis}
After identifying the neutral regions in the 21-cm observations, we need to summarize the information contained in them. In this study, we will use two methods to quantify the sizes of the identified structures, namely the mean-free-path (MFP) and friends-of-friends (FOF) size distributions. In addition to the size distributions, we also use the Euler characteristic to quantify the topology of the structures. In this section, we describe all these methods.

\subsubsection{MFP size distribution}
\citet{2007ApJ...669..663M} proposed a ray tracing method to determine the size distribution of the region of interest in the data cube. In our binary field, we randomly select a point in our region of interest (say the neutral regions) and shoot a ray in a random direction. The ray is stopped when it reaches the boundary of the region of interest, at which point the length of the ray is recorded. After repeating this this process a sufficient number of times, we will have obtained a representative sample of the sizes ($R$) present in our binary field which we present using the PDFs of the lengths of the rays. Throughout this work, we use $10^7$ number of rays to estimate the size distributions. When applied to ionized regions the size distribution is usually referred to as the bubble size distribution (BSD). When we apply it to neutral regions we will use the terminology island size distribution (ISD).

\subsubsection{FOF size distribution}
\label{sec:fof_method}
The FOF method is based on the idea of finding clustering patterns in data sets \citep[e.g.,][]{ivezic2014statistics}. Any two points belong to the same cluster when the distance between them is smaller than a chosen linking length. The Hoshen-Kopelman algorithm is an efficient way to define such clusters in gridded data \citep{Hoshen1976PercolationAlgorithm}. The FOF method was introduced by \citet{Iliev2006SimulatingReionization} to determine the volume ($V$) of the ionized regions in gridded x$_\mathrm{HII}$ data from an EoR simulation. The information is presented in a histogram of cluster volumes, which we will call the FOF-BSD.


During the middle and late stages of reionization, the FOF-BSDs show a characteristic bimodal distribution in which one large cluster contains most of the ionized cells and the rest are distributed over a range of small clusters \citep{Friedrich2011TopologyReionization,Furlanetto2016ReionizationTheory}. The FOF-BSDs always show a clear gap between these small clusters and the large cluster. \citet{Furlanetto2016ReionizationTheory} pointed out that this behaviour is typical for a percolation process and that the single large cluster in the FOF-BSD is the so-called percolation cluster, which contains most of the total ionized volume after the Universe has approximately reached 10 per cent ionization.

With its focus on connectivity, the FOF-BSD/ISD contains very different information from the MFP-BSD/ISD although both in some sense characterise the sizes of the chosen region of interest. The FOF-BSD/ISD allows us to study the growth of the percolation cluster together with the size distribution of clusters that are not yet part of the percolation cluster. The MFP-BSD/ISD gives the distribution of typical sizes which apart from sample variance will be identical for three-dimensional volumes or two-dimensional slices. For more discussion about differences in the information extracted by the two methods, see \citet{Giri2018BubbleTomography}.

\subsubsection{Euler characteristic}
\label{sec:euler_char}
Topology studies the properties of space that are preserved under continuous deformations, such as stretching, crumpling and bending, but not tearing or gluing. The important topological properties include connectedness and compactness. In this study, we use the $\chi$ to describe the topology of structures. $\chi$ is a topological invariant that is proportional to the integral of the Gaussian curvature over the surface of the structure and is also known as the third Minskowski functional (see table 1 in \citet{1996dmu..conf..281S} for the exact conversion factor). 
For any manifold $M$, the $\chi$ is given as follows
\begin{eqnarray}
\chi (M) = \#\mathrm{Components}(M) - \#\mathrm{Tunnels}(M) + \#\mathrm{Cavities}(M) \ .
\end{eqnarray}

In our case, the data set represents the manifold in which the ionized and neutral regions form structures, which are very complicated. Therefore we must triangulate the structure with well defined geometrical components in such a way that the topological properties of the structure is retained.
A cubical complex is one such triangulation method, which is composed of points, line segments, squares and cubes.
We explain the conversion of our data set into a cubical complex in Appendix~\ref{sec:cubical_complex}. 
Once our data set is converted to a cubical complex, the determination of topological measures is simple.
We have implemented the algorithm given by \citet{Gonzalez-Lorenzo2016FastComplexes} to calculate $\chi$ in the python package {\sc tools21cm}\footnote{A python package to analyse the 21-cm signal and extract information from it. It can be downloaded from \url{https://github.com/sambit-giri/tools21cm}}. A brief description of the algorithm is given in Appendix~\ref{sec:calc_chi}. For a Gaussian field, the $\chi$ has a specific evolution \citep[see fig.~1 of][]{Schmalzing1997BeyondStructure}. We use this property to validate our implementation of the $\chi$ algorithm (see Appendix~\ref{sec:calc_chi}).

The $\chi$ has been used extensively to understand the topology of large scale structures \citep[e.g.,][]{Gott1986TheUniverse,Mecke1993RobustUniverse,Schmalzing1997BeyondStructure}. More recently, $\chi$ (or the genus $g = 1-\chi$) was used to describe the topology of reionization \citep[e.g.,][]{Gleser2006TheFunctionals, Lee2008TheReionization, Friedrich2011TopologyReionization}. The first two papers studied the $g$ of the neutral density field, whereas the last one studied the topology of the (binary) ionization field. \citet{Yoshiura2017StudyingReionization} studied the $\chi$ of the 21-cm signal and the contributions from the neutral hydrogen field, the density field and the spin temperature. Below we will consider the evolution of $\chi$ of the neutral field as identified by our superpixels method from a simulated noisy 21-cm data set.

\begin{figure*}
  \centering
  \includegraphics[width=1.0\textwidth]{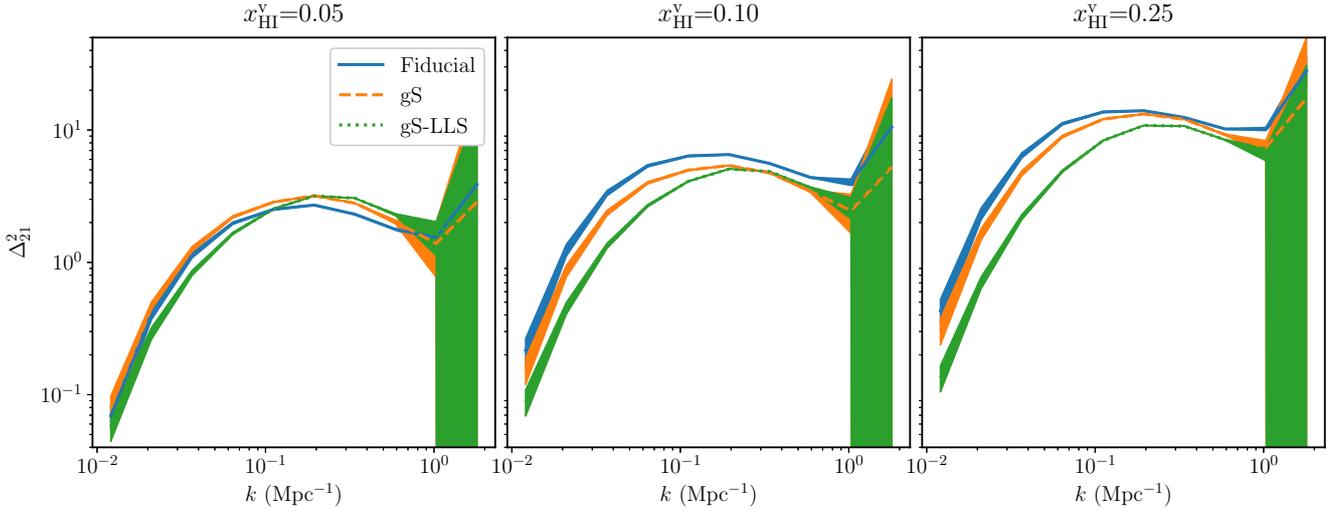}
  \caption{The 21-cm power spectra during the late stages of reionization for the three scenarios from Table~1. {The shaded regions are uncertainties on the power spectra due to sample variance and the system noise associated with a 1000 hour observation with SKA-Low.} These power spectra all have similar shapes, which will make it difficult to distinguish both the evolution and the different scenarios.}
  \label{fig:simframe_ps_compare}
\end{figure*}

\begin{figure*}
  \centering
  \textbf{\LARGE Early stages \qquad \qquad \qquad Late stages \qquad \qquad}\par\medskip
  \includegraphics[width=1.0\textwidth]{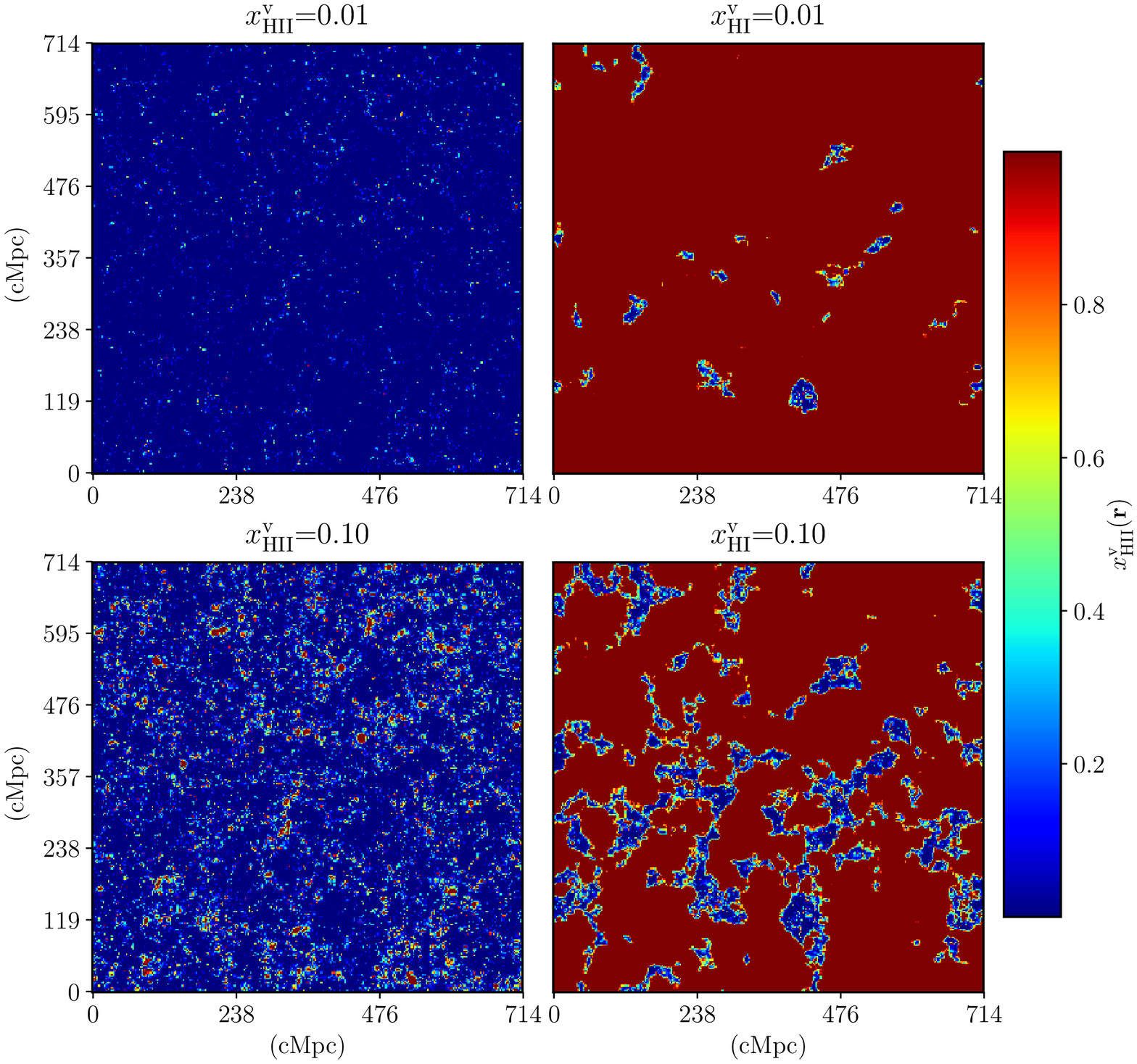}
  \caption{The ionized and neutral regions in our Fiducial simulation. The top panels shows the slices when the simulation box is 1 per cent ionized and neutral in left and right panels respectively. Visually the ionized regions appear different from the neutral regions. The neutral regions are larger and therefore fewer in number compared to the ionized regions. The lower panels present similar visual contrast when 10 per cent of the simulation box is ionized and neutral in left and right panels respectively.}
  \label{fig:mock_slices}
\end{figure*}

\section{Results}

\subsection{Power spectra}
The 21-cm power spectrum is a useful statistical probe of the fluctuations in the signal and has been the main metric used for characterizing the signal. In the high $T_\mathrm{s}$ limit, the 21-cm power spectra from before reionization starts is defined by the matter density fluctuations. Once reionization begins, the power spectrum is modulated by the growth of ionized bubbles \citep[e.g.,][]{Lidz2008DetectingArray} and the density fluctuations inside the ionized bubbles no longer contribute to the 21-cm signal. As a result, the 21-cm power spectrum becomes insensitive to the fluctuations below the characteristic bubble scale.

In Fig.~\ref{fig:simframe_ps_compare} we show the 21-cm power spectra during the very late stages of reionization for the three different simulations described in Section~\ref{sec:sims}. {The shaded regions indicate the uncertainty due to sample variance and noise for a 1000 hour observation with SKA-Low. The uncertainty due to system noise is given by the noise residuals after subtraction of the (average) noise power spectrum. We calculate these using the procedure described in section~3.2.1 of \citet{Jensen2013ProbingDistortions}.}

The power spectra of the three simulations are similar in shape with small differences in amplitude and all three show only minor evolution of the shape. The absolute magnitude of the 21-cm power spectra derived from observations may be inaccurate due to the presence of residual foreground signals. Differences in the shape of the 21-cm power spectra will then make it easier to distinguish between different reionization scenarios. These rather isomorphic 21-cm power spectra from the late phases of reionization thus contain little information both in terms of evolution and in terms of source properties. This motivates our current study to investigate whether other metrics, such as the BSDs and ISDs, may be more useful for characterising the process during these late stages.

\subsection{Comparison with ionized regions}


\begin{figure*}
  \centering
  \includegraphics[width=1.0\textwidth]{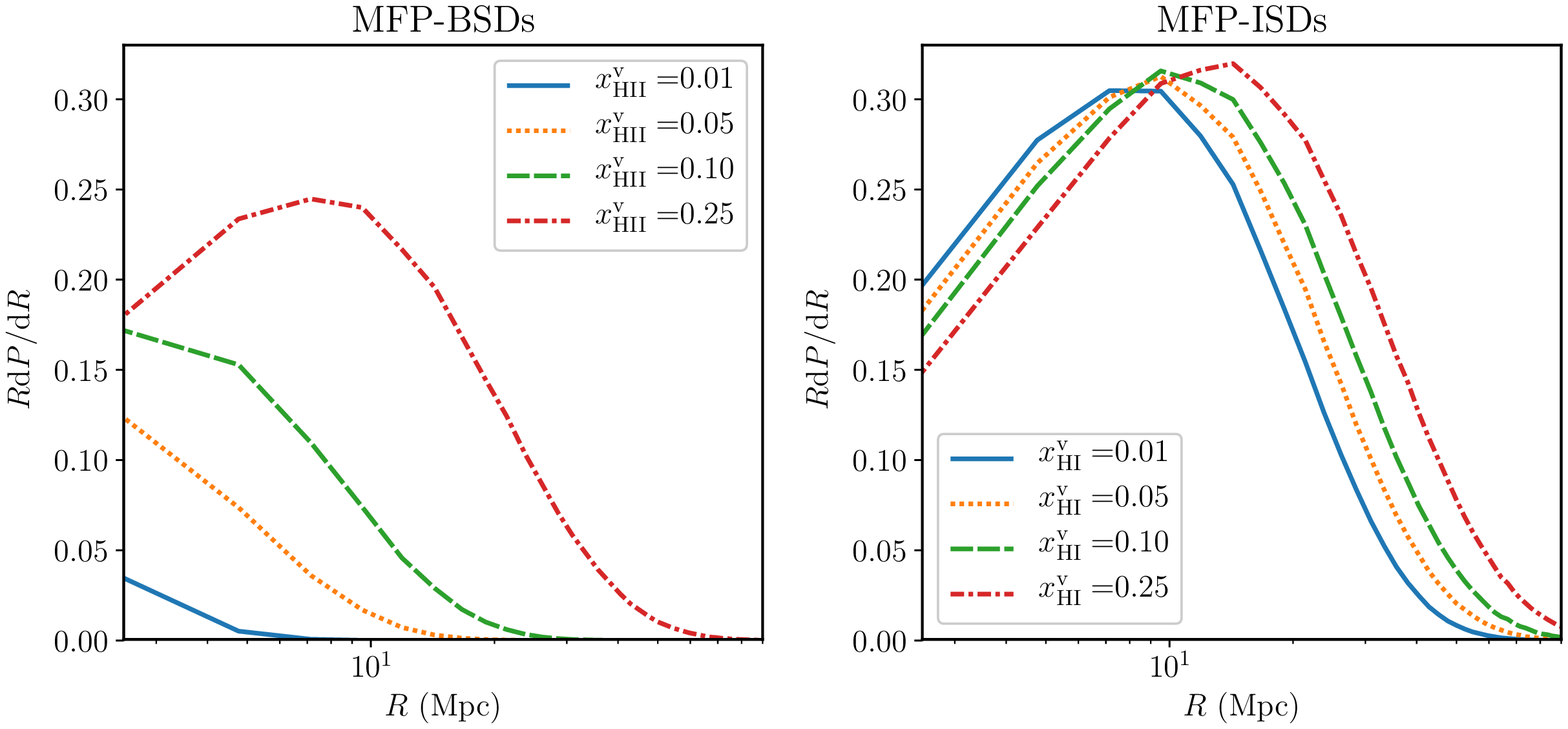}
  \caption{The size distribution of the ionized bubbles (left panel) and neutral islands (right panel) estimated using the MFP method in our Fiducial simulation. The MFP-BSDs and MFP-ISDs are calculated from the simulation boxes that have same volume filling. The corresponding volume filling fraction is mentioned in the legend. The evolution of the sizes of the ionized bubbles are faster compared to that of the neutral islands.}
  \label{fig:simframe_mfp_bsd_compare}
\end{figure*}

\begin{figure*}
  \centering
  \includegraphics[width=1.0\textwidth]{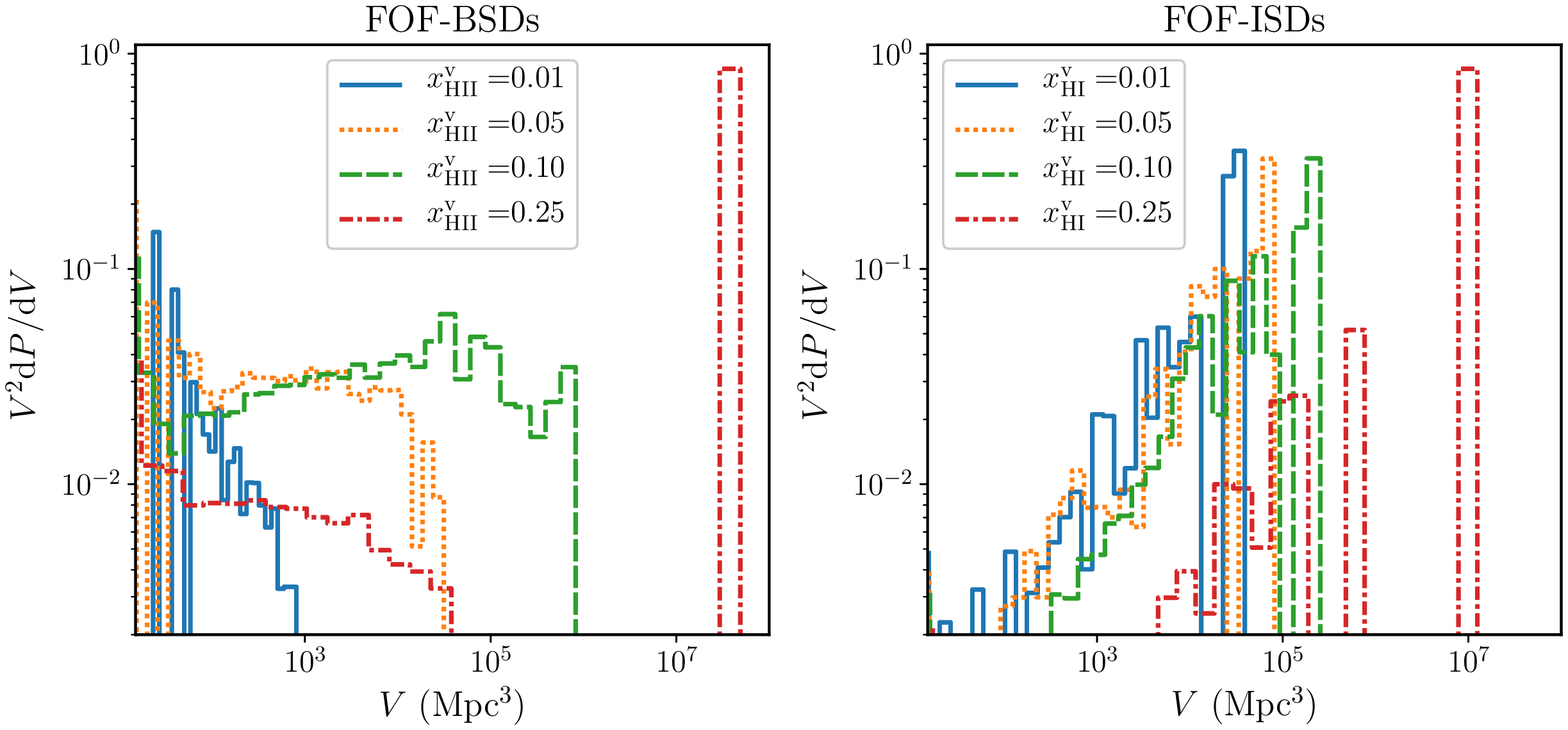}
  \caption{Same as Fig.~\ref{fig:simframe_mfp_bsd_compare} but with the sizes determined using the FOF method. We observe the null gap, which separates the two population of sizes, in the size distribution of both ionized bubbles and neutral islands. The null gap is a characteristic signature of the formation of the percolation cluster.}
  \label{fig:simframe_fof_bsd_compare}
\end{figure*}

We start by highlighting the differences between the evolution of ionized bubbles and neutral islands in our simulation data. When comparing their shapes, sizes and distribution at equivalent stages of their evolution, we find them to be completely different. Fig.~\ref{fig:mock_slices} gives a visual impression of the ionized bubbles and neutral islands when both have similar volume filling fractions. The ionized regions during the beginning of reionization ($x_\mathrm{HII}^\mathrm{v}$ = 0.01, 0.10) are mostly small and isolated structures in the entire volume because the ionized bubbles are located at the positions of the ionizing sources. However, the neutral regions during the end of reionization ($x_\mathrm{HI}^\mathrm{v}$ = 0.01, 0.10) are much larger structures of which many fewer exist. They trace the cosmic voids in which no or few sources of ionizing photons exist.

Although the details will depend on the actual properties of the sources of ionizing photons, this result will be true for any reionization process in which the sources are associated with dark matter haloes, which trace the underlying density field and the source population grows exponentially. In such a scenario, reionization will be `inside-out' in nature, meaning that higher density regions will reionize first and the mass weighted mean ionization fraction will be larger than the volume weighted one.

In the subsequent subsections, we will compare the differences between the ionized and neutral regions using the statistical probes presented in Section~\ref{sec:structure_analysis}. {Even though we present our results in this section using the `Fiducial' simulation, we find similar trends in all our simulations.} 




\subsubsection{Size statistics}

In order to understand the statistical differences between the ionized bubbles and the neutral islands, we look at the respective size distributions using both the MFP and FOF methods. As explained in Section~\ref{sec:fof_method}, these two distributions measure different size properties of the regions they are applied to. 

In Fig.~\ref{fig:simframe_mfp_bsd_compare}, we show the evolution of the MFP BSDs (left panel) and ISDs (right panel) for similar volume filling fractions of ionized and neutral regions respectively. We see that at any given filling factor the peak of the ISD is at larger sizes than for the BSDs and tail of the distribution also extends to larger sizes. Furthermore, the ISDs evolve only slowly with filling factor whereas the BSDs show a very strong evolution \citep[see also][]{2007ApJ...669..663M}. As we saw above in Fig.~\ref{fig:mock_slices}, the neutral islands are a set of relatively few but large regions. Therefore a small growth in the sizes of the neutral region can cause an appreciable level of increase in volume filling. However, our results do not agree with the finding in \citet{Xu2017IslandFAST:Reionization} where the peak position of the ISDs stays unchanged during reionization. Although the evolution is not strong, the peak of the ISD from our simulations moves to larger sizes with increasing $x_\mathrm{HI}^\mathrm{v}$. {For a different choice of the definition of neutral regions, \citet{Xu2017IslandFAST:Reionization} do find some evolution of the peak position, see their fig.~5. We plan to investigate these differences in future work.}

Fig.~\ref{fig:simframe_fof_bsd_compare} shows the same evolution as seen in the FOF BSDs and ISDs. Here, we plot $V^2 \frac{\mathrm{d}P}{\mathrm{d}V}$ instead of $V \frac{\mathrm{d}P}{\mathrm{d}V}$ because most the volume is occupied by the percolation cluster and its importance would not be visible in a plot of $V \frac{\mathrm{d}P}{\mathrm{d}V}$ \citep{Lin2016TheReionization}. We see that both the BSD and ISD develop a single dominating percolation cluster for high enough volume filling fractions. However, in everything else the ISDs and BSDs are very different.

In the FOF-BSD, the distribution of the smaller bubbles is relatively flat, except at the earliest stages, $x_\mathrm{HII}^\mathrm{v}=0.01$. This means that all sizes contribute equally to total filling factor or that $\mathrm{d}n/\mathrm{d}V\propto V^\tau$ with $\tau\approx -2$. \citet{Furlanetto2016ReionizationTheory} showed that this is expected from percolation theory if there is some clustering of the sources as for a random field the exponent would be $\tau=-2.18$. However, in the ISDs we note that the distribution of smaller regions shows a strong positive slope, meaning that the larger regions dominate over the smaller ones. This implies that the clustering of the neutral islands is much stronger as the exponent we derive $\tau\approx -1.5$ deviates strongly from -2.18. Furthermore, even though the neutral islands correspond to low density voids, the value is also higher than the one derived by \citet{2006MNRAS.367.1629S} for voids in N-body simulations, $\tau\approx -1.95$.


\subsubsection{Percolation curve}
\label{sec:reslt_perc}

\begin{figure}
  \centering
  \includegraphics[width=0.46\textwidth]{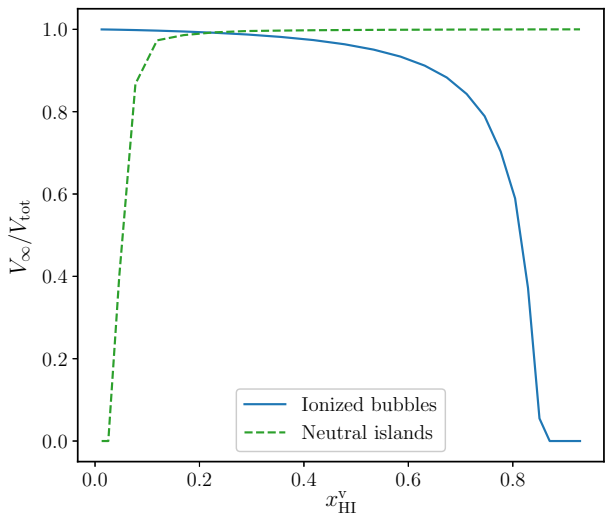}
  \caption{The percolation curve of the ionized and neutral regions in the Fiducial simulation. The ordinate gives the ratio of the volume of the percolation cluster {($V_\mathrm{\infty}$)} to the total volume {($V_\mathrm{tot}$)} of the respective region, ionized or neutral. The abscissa shows the reionization epoch with the average neutral fraction ($x^\mathrm{v}_\mathrm{HI}$) of the simulation box.}
  \label{fig:perc_curves}
\end{figure}

\citet{Furlanetto2016ReionizationTheory} showed that reionization behaves as a percolation process. After a certain threshold ($x_\mathrm{HII,c}^\mathrm{v}$), an infinite cluster of ionized regions is formed which connects more than 50 per cent of the total ionized region. They define the infinite (or percolation) cluster as the ionized region that spans the opposite faces of the data set in at least one direction. Under periodic boundary conditions this makes it infinite in size, hence the name. They found that percolation happens at a volume filling {by ionized regions} of approximately 10 per cent in their simulation volume. \citet{Giri2018BubbleTomography} found the same behaviour for the simulated 21-cm signal observations.
 
In this section we compare the percolation properties of the ionized and neutral regions. We can study the evolution of the percolation cluster using the percolation curve which shows the evolution of the fraction of the total neutral volume contained in the neutral percolation cluster.
Fig.~\ref{fig:perc_curves} shows the percolation curves for both the ionized and neutral regions. We find that also in this simulation the ionized regions percolate at $x_\mathrm{HII,c}^\mathrm{v}\approx 0.1$. After the percolation threshold, the growth of the percolation cluster depends on the size and distribution of the isolated regions \citep[e.g.,][]{Strelniker2009FractalsPercolation, Saberi2015RecentApplications}. We notice that the percolation curve for the ionized bubbles only slowly approaches a value of 1 after the formation of percolation cluster. This is indicative of the isolated ionized bubbles being scattered and small. 

We find the percolation threshold of the growing neutral islands to be substantially lower at $x_\mathrm{HI,c}^\mathrm{v}\approx 0.05$. This value is less accurate as the time sampling of the reionization process is insufficient to capture the fast changes occurring during the end of reionization. However, the percolation threshold is definitely lower than for the ionized regions. We also find that the percolation curve reaches one faster compared to the one for ionized regions. This can again be understood from the larger sizes of the neutral islands: isolated neutral regions connect to the percolation cluster faster. \citet{Furlanetto2016ReionizationTheory} reported similar findings about the percolation of the neutral islands. A detailed percolation study of the neutral islands is beyond the scope of this paper.

\subsection{Neutral islands in simulated observations}
Having studied the power spectra, size distributions and percolation in the simulation results, we now turn our attention to analysing simulated observations. We produce a simulated noisy 21-cm signal data cube according to the procedure outlined in Section~\ref{sec:observed_signal} and process it with the structure identification process described in Section~\ref{sec:method_ident} to identify neutral regions. Using this we derive size distributions and the $\chi$ of the neutral islands. Here we use the Fiducial simulations to illustrate the impact of telescope limitations on the statistics derived from observations.

\subsubsection{Size statistics}

\begin{figure*}
  \centering
  \includegraphics[width=1.0\textwidth]{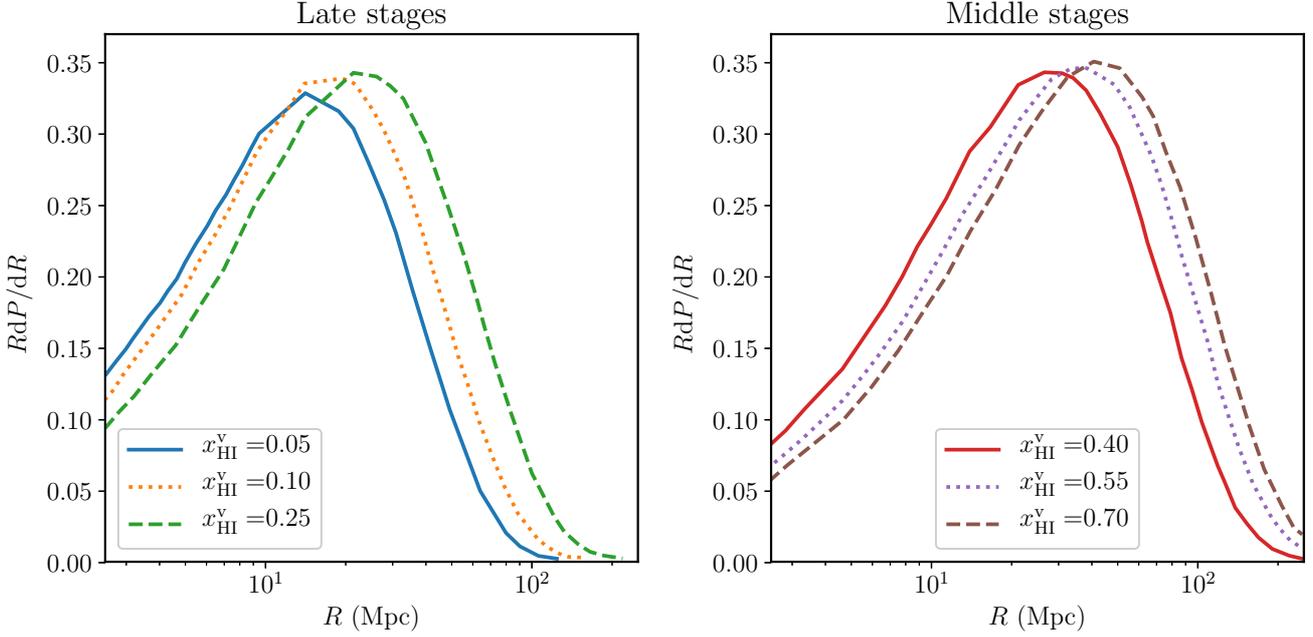}
  \caption{The MFP-ISDs for neutral islands identified by the superpixels method from the simulated 21-cm observations during the late (left panel) and middle (right panel) stages of reionization. We see the evolution of the ISDs for the different epochs, represented by the average neutral fraction $x^\mathrm{v}_\mathrm{HI}$.}
  \label{fig:obsframe_mfp_bsd_compare}
\end{figure*}

In \citet{Giri2018BubbleTomography} we studied the difference between observed and intrinsic BSDs and found them to be substantially different. The resolution limit of the observations makes it impossible to identify the smaller ionized regions which affects the BSDs in two ways. Firstly, the amplitude of the peak of the (normalized) distribution is larger for the observed BSD as the contributions from the smallest regions disappears. Secondly, the evolution of the peak value is much slower in the observed BSDs than in the intrinsic BSDs as the intrinsic distribution for a long time peaks below or around the resolution limit.

We show the evolution of the MFP-ISDs of the identified neutral islands during the EoR in Fig.~\ref{fig:obsframe_mfp_bsd_compare}. While comparing these to the ISDs obtained at the simulation resolution, we see that they are not all that different. Both show a relatively slow evolution of the peak position and the peak amplitudes are quite similar. Clearly the effects introduced by the telescope do not affect the ISDs as seriously as they affect the BSDs.

The main reason for this is that the distribution of neutral regions is strongly skewed towards larger sizes, which means that the lower resolution of the observations does not impact the derived ISDs very much as most of the neutral regions are easily resolved. An additional factor which helps is that the noise levels are lower at higher frequencies which makes it easier to identify reliably the neutral regions.

\subsubsection{Euler characteristics}
\label{sec:reslt_obs_chi}


\begin{figure}
  \centering
  \includegraphics[width=0.46\textwidth]{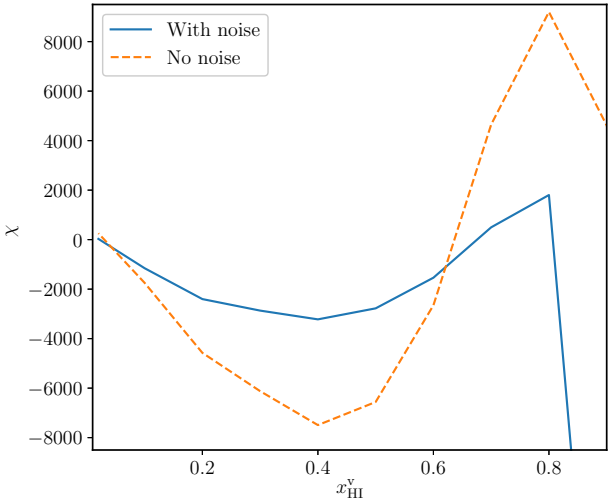}
  \caption{The Euler characteristics ($\chi$) calculated at various stages of reionization as characterised by $x^\mathrm{v}_\mathrm{HI}$. The solid (blue) curve and the dotted (green) curve show $\chi$ for mock observations with and without noise respectively. 
  Both curves show a similar behaviour with an early rise of $\chi$ followed by a sharp decline into negative values with a subsequent increase although positive values are never again reached for the observation with noise. Negative values indicate that tunnels dominate over isolated cavities or islands.}
  \label{fig:sim_chi}
\end{figure}

Next we consider the topological measure of the $\chi$. We calculate this quantity using the method described in Section~\ref{sec:euler_char} and Appendices~\ref{sec:cubical_complex} and \ref{sec:calc_chi}. The $\chi$ is an invariant which means that for a binary field the topology is the same for both components. We will describe the results in terms of the neutral regions as the structure. See \citet{Friedrich2011TopologyReionization} for a description in terms of ionized bubbles.


Fig.~\ref{fig:sim_chi} shows the value of $\chi$ at various epochs in our Fiducial simulation. We show two curves, one for noiseless mock observations and the other for mock observations with noise. 
We see that noise reduces the magnitude of $\chi$ over the whole history of reionization. The structure identification method will find less features due to presence of noise. Therefore the number of components, tunnels and cavities found reduces. Even though the magnitude of $\chi$ goes down due to noise, both curves still show qualitatively similar behaviour. When including telescope noise the $\chi$ values estimated for the early stages of reionization ($x^\mathrm{v}_\mathrm{HI} \gtrsim 0.8$) deviate strongly from the noiseless values. The ionized regions identified at these high $z$ are unreliable as the structure identification method struggles due to the higher noise levels and the smaller ionized regions \citep[see fig.~9 of][]{Giri2018OptimalObservations}. 

During early stages of reionization, we see a rise in the value of $\chi$ as more and more ionized regions appear as cavities in the neutral Universe. Once the neutral fraction has reached 80 per cent, the percolation cluster of ionized regions has formed which means that many of the ionized regions are now connected to each other. At the same time the contribution of tunnels through the neutral medium is starting to become important. As tunnels reduce the value of $\chi$, the curve reaches a maximum. Around $x^\mathrm{v}_\mathrm{HI}\approx 0.6$ the contribution of tunnels starts to dominate over the isolated ionized regions and the $\chi$ becomes negative. As the remaining neutral regions become more and more island-like, the value of $\chi$ starts to increase again. However, the $\chi$ for the observations with noise never reaches positive values which implies that up to the end of reionization the neutral islands will appear to be perforated by ionized tunnels.

These results once more illustrate the asymmetry between the morphology of the early ionized regions and the late neutral islands. The former constitute a large set of many isolated regions whereas the latter form a smaller set of larger regions with a high degree of porosity.


\subsection{Different reionization scenarios}

\begin{figure*}
  \centering
  \includegraphics[width=1.0\textwidth]{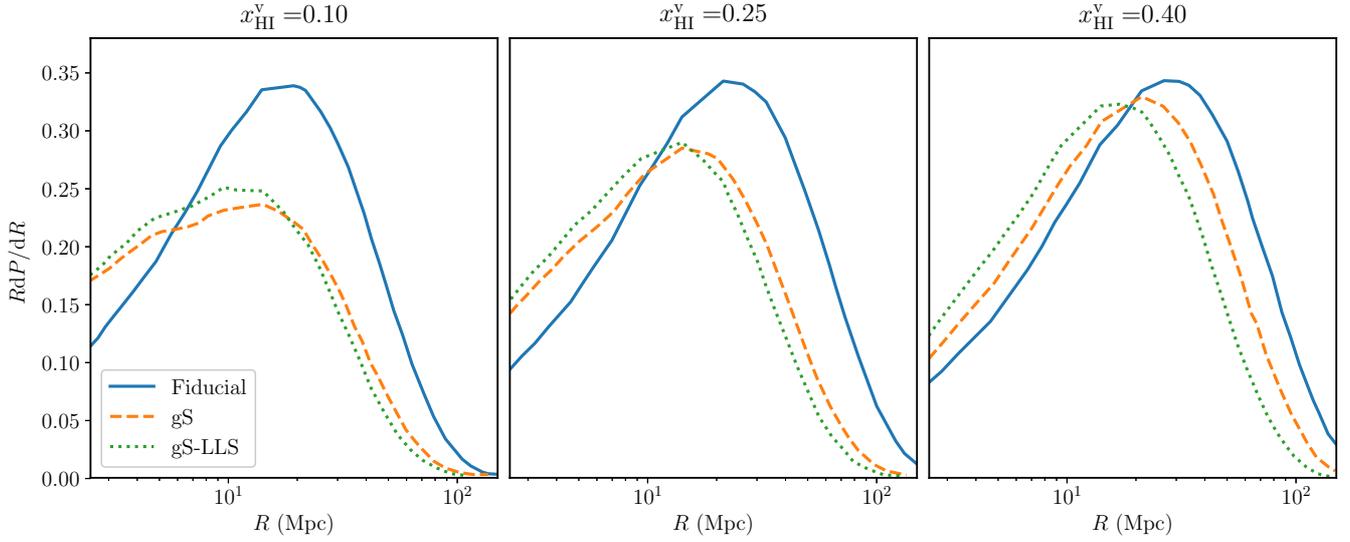}
  \caption{The MFP-ISDs from our set of three simulations. The mean neutral fraction increases from left to right. We see that the Fiducial simulation in which reionization is caused by more massive haloes has larger neutral islands at all times. The ISDs of two simulations in which lower mass haloes also contribute peak at smaller values. Including position- and redshift-dependent LLS pushes the ISDs to even smaller values although the difference is relatively minor.}
  \label{fig:obsframe_mfp_bsd_compare_model}
\end{figure*}

\begin{figure*}
  \centering
  \includegraphics[width=0.85\textwidth]{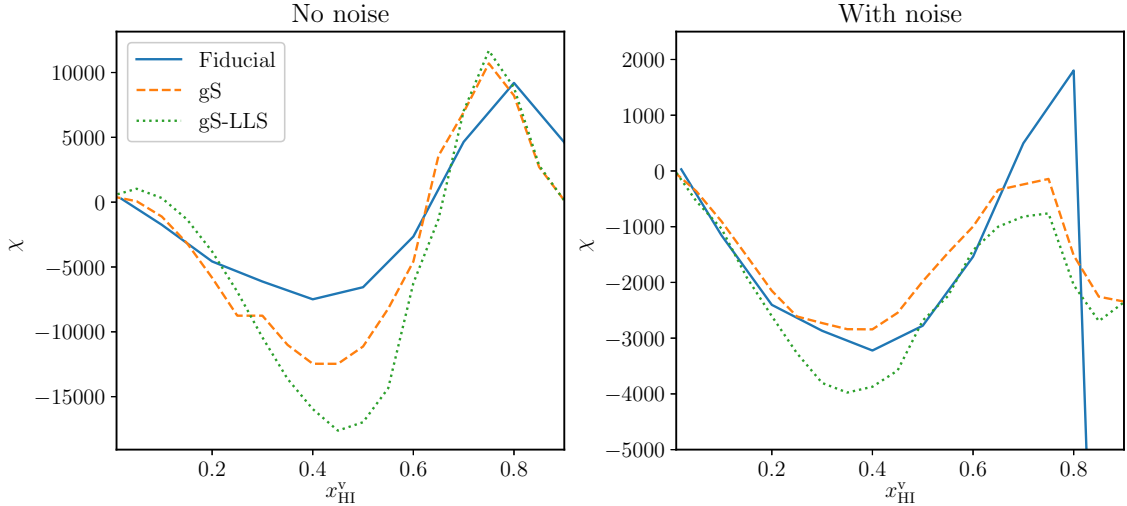}
  \caption{The evolution of the Euler characteristic ($\chi$) of the neutral islands for the three simulations. We show the $\chi$ for the simulated 21-cm observations without and with thermal noise in the left and right panel respectively. 
  Noise affects the earliest phases more as the superpixels method confuses the ionized regions with noise.}
  \label{fig:obsframe_chi_compare_model}
\end{figure*}

All the results above were obtained for one reionization simulation which assumes specific properties for the sources of reionization. One might wonder how different the results are for different assumptions about the sources. A full exploration is beyond the scope of this paper but we will consider one other scenario for the sources. A further question we want to address here is how dependent the properties of neutral islands are on the choice of subgrid model for the LLS. As pointed out in the Section~\ref{sec:intro}, the presence of LLS is an important factor at the end of reionization.

The large body of work on ionized regions has shown that details of the size and shape evolution of those regions are sensitive to the source properties. This means that 21-cm observations can help in constraining the properties of early galaxies. However, the BSDs towards the end of reionization start losing their meaning and ISDs become a better metric for describing the state of the IGM. In this section we compare ISDs from mock observations derived from three different reionization simulations (see Section~\ref{sec:sims}). We also consider the evolution of the $\chi$ in these mock observations. 




In Fig.~\ref{fig:obsframe_mfp_bsd_compare_model} we plot the MFP-ISDs for three different reionization scenarios at different reionization epochs. These MFP-ISDs are constructed from the noisy simulated 21-cm signal after identifying the neutral islands with the superpixel method. We compare the ISDs at similar intrinsic global neutral fraction. The additional low mass sources present in the `gS' and `gS-LLS' cases shifts the entire reionization history to larger redshifts. However, we will compare the results at similar global neutral fractions as the differences will help us break the degeneracy when the reionzation history would be the same.

Before addressing the differences we first want to point out that all three models show the same qualitative behaviour as our Fiducial scenario regarding the evolution of late neutral islands. The distribution is dominated by relatively large regions, percolation is relatively late and SKA1-Low observations resolve most of the neutral islands. We therefore expect these results to be robust even if the final stages of reionization are not well understood.

We see that the in the case of `gS' the ISDs at all phases of reionization is shifted to smaller sizes. There are fewer very large neutral islands and more small ones and the peak occurs at smaller sizes. For the later stages the ISD is also wider, mostly because more smaller islands are present. Clearly a different population of sources affects the ISDs. At the same time we see that a different LLS model only has a minor impact on the ISDs as the curves for `gS' and `gS-LLS' are quite similar, although the latter shows a shift to even smaller sizes. However, clearly the details of how the opacity of LLS are implemented only has a minor effect of the ISDs compared to the impact of a different source population.

At the same global neutral fraction, the gS simulation also shows a BSD that is shifted to smaller values compared to the Fiducial case. The less clustered low mass sources lead to a more fragmented distribution of smaller ionized regions. However, if the ionized regions are smaller, should the neutral regions not be larger instead of smaller? The answer to this apparent paradox lies in the mathematical `packing problem'. As we explain and illustrate with a toy model in Appendix~\ref{sec:toy_model}, packing a volume with smaller spheres also fragments the area outside the spheres into smaller units. So a more fragmented reionization also leads to smaller neutral islands. The position and time-dependent LLS model leads to further fragmentation of ionized regions and thus a further shift to smaller sizes for the ISDs.

All our scenarios show that until quite late (90 -- 95 per cent ionization) the longest line of sight through a neutral island can be more than 100~comoving Mpc and that the frequency of such lines of sight is at least one per cent. These sizes are reminiscent of the long absorption trough at $z=5.6$--5.8 in the spectrum of the quasar ULAS~J0148+0600 reported by \citet{2015MNRAS.447.3402B}. Our models clearly produce substantial variations in the ultraviolet background (UVB) until quite late. However, we have not run our models until those late redshifts and as we saw in Section~\ref{sec:sims}, our models do not reproduce the correct transition to the post-reionization state of the IGM. Therefore the final word on this requires applying carefully calibrated models such as the ones in \citet{Kulkarni2019Large5.5} and \citet{Keating2019LongHydrogen} to volumes equivalent to ones we study here.

Finally we also compare the evolution of the Euler characteristic $\chi$ for these three cases of mock observations. Fig.~\ref{fig:obsframe_chi_compare_model} shows the evolution of $\chi$ against the global neutral fraction ($x^\mathrm{v}_\mathrm{HI}$) for observations without and with thermal noise in the left and right panel respectively. {Here the gS and gS-LLS cases use the same resolution and noise as the Fiducial case. This allows us to compare the results on an equal footing.} 

{Considering the noise-less results first, we see that all three models show similar topologies at late times, approaching zero below, although the gs-LLS case actually becomes slightly positive before reaching zero. The right panel shows that these differences are wiped out by the noise. Around the midpoint of reionization, the topologies differ more, with the gS-LLS case reaching the most negative values. The Euler characteristic can only become negative due to tunnels, so this shows that the gS-LLS case has considerably more tunnels than the gS case which in its turn has more than the Fiducial case. Again these differences become a lot less when noise is added, although the gs-LLS case still reaches the lowest values. Even earlier, all three cases show a positive maximum with the gS and gS-LLS cases showing very similar results. In the presence of noise these results are drastically changed as the maxima for the gS and gS-LLS cases do not even reach positive numbers. 

All of this can be understood from the fact that the gS cases contain many more small \HII} {regions due to the presence of low mass sources and that the presence of the LLS inhibits the growth of the regions which leaves them more fragmented. When adding noise the small regions are affected more than the larger ones which explains the large difference with the Fiducial case.}

This shows that even when ISDs are quite similar, the topologies can differ and vice-versa (cf.~Fig.~\ref{fig:obsframe_mfp_bsd_compare_model}). Clearly there is additional information present in topological measurements.



\section{Summary and discussion}
\label{sec:conclusions}

In this paper, we present a study of the neutral islands which are the last regions to reionize. We consider their general properties and compare these to those of the first ionized bubbles to form. We also investigate how well these properties can be measured considering the limited resolution and expected noise levels of imaging observations with the future SKA1-Low. Although properties of neutral islands occasionally featured in previous works which mainly focused on the ionized bubbles \citep[e.g.,][]{2007ApJ...669..663M, Iliev2014SimulatingEnough}, we here present a 
systematic exploration of the properties of late neutral islands based on the results of fully numerical simulations.

Our results strongly support dedicating observing time to the later stages of reionization. Even at global ionization fractions of more than 95 per cent, the remaining neutral islands are excellent observing targets for tomographic imaging observations, as they until very late remain relatively large, and therefore can be resolved, and the problems of both foregrounds and instrumental noise are less severe at lower redshifts. At these late times the 21-cm power spectra are not very sensitive to the source properties and do not show much evolution except for their amplitude. We find that the size distributions of late neutral islands evolve and depend on the source properties and thus contains useful information about the reionization process. Tomographic obervations which can identify the late neutral islands are thus to be preferred for late reionization.

The appearance of the first ionized bubbles and the disappearance of the last neutral islands proceed very differently from each other. The former process is dominated by many small bubbles, whose size distribution evolves rapidly. Even after percolation when larger ionized bubbles have formed, the smaller bubbles keep contributing substantially as the FOF-BSD shows. The late neutral islands are instead dominated by the larger structures.
The size distribution of neutral islands evolves slowly, but, in contrast to what was found by \citet{Xu2017IslandFAST:Reionization}, does evolve to smaller sizes as reionization progresses.

{We find that at least one per cent of the lines of sight through a neutral region is longer than 100 comoving Mpc until very late stages (90-95 per cent reionized). If this stage of reionization is only reached by redshift 5.5 or later, as suggested by \citet{Keating2019LongHydrogen}, our results suggest that the long absorption trough seen in the spectrum of ULAS J0148+0600 could easily be caused by such a late neutral island.}

In terms of topology, both the ionized bubbles and neutral islands percolate at some point, however, the islands percolate much more rapidly. Already at a global neutral fraction of $\sim 5$ per cent, most of the still neutral IGM is contained within one large connected region, whereas it requires a global ionized fraction of $\sim 10$ per cent to achieve the same for the ionized bubbles. At the same time the values of the $\chi$ suggests that the remaining neutral islands are porous and contain many tunnels as the values remain negative or close to zero. The values for the early stages of reionization are positive and large which implies that the topology is dominated by isolated bubbles.

Reionization scenarios in which lower mass sources contribute are obviously characterised by ionized bubble size distributions which peak at smaller bubble sizes. We find that also the ISDs shift to smaller sizes. This perhaps counter-intuitive result can be understood in terms of the mathematical packing problem: smaller objects cover space more efficiently and thus also leave smaller gaps between them. The conclusion is therefore that in general scenarios with different BSDs will also have different ISDs.

We do not present an extensive overview for many different reionization scenarios which will obviously be required for establishing to which extent the late neutral island properties can be used to distinguish different source models. Such an investigation would probably be best done in the context of a parameter study in which the ISDs are used as a metric, either in a foreward \citep{Greig201521CMMC:Signal} or backward \citep{Shimabukuro2017AnalysingNetworks} modelling approach. Such a study would immediately establish for which parts of parameter space the neutral island statistics are particularly distinguishing. 

In addition, given the existing measurements of the ionization state of the post-reionization Universe and the measurements of the UV background around $z=6$, such a study should also take into account these constraints. Future measurements of the properties of late neutral islands combined with Ly$\alpha$ forest measurements of the post-reionization IGM could then lead to a much improvement understanding of galaxy evolution during this transition phase.

An aspect which requires more investigation is the effect of spatial resolution on the results presented here. Although our simulations have enough dynamic range to show conclusively that larger neutral islands are more common than smaller ones, there will likely be a population of neutral regions much below our resolution limit, namely contained inside collapsed objects or their environments. It is well-known that after reionization the remaining neutral hydrogen is found mostly in Damped Lyman-$\alpha$ (DLA) systems and constitutes some 3 per cent of the total hydrogen in Universe \citep{2015MNRAS.452..217C}. Our simulations do not contain this component and when we quote numbers such as 1 per cent neutral this should be understood to be valid for the large scale IGM. These DLAs would formally constitute a population of very small neutral islands, at $\sim 50$~kpc some 10 times smaller than our resolution. Not only do they differ in size, they also have a different physical cause. Whereas the large neutral islands are neutral due to a lack of nearby sources of ionizing photons and are therefore associated with low density areas in the IGM, the DLAs are neutral because of their high density.

However, the presence of small scale neutral structures does have an effect on the large neutral islands in the IGM namely because they limit the mean free path of ionizing photons, thus reducing the number of sources which can photoionize the islands. We include this effect as a subgrid model in our of our scenarios and find that their presence shifts the ISD to smaller values for the same global neutral fraction. This is another application of the packing problem described above. However, we find the impact of the small scale absorbers on the ISD to be small and not to change our conclusions qualitatively.

Late neutral islands will form a perfect target for observations with the future SKA1-Low telescope. We showed that for deep exposures of 1000~h, they will be easily identified and because of their large sizes, their measured size distribution will be similar to the intrinsic one. This in contrast with the measured BSDs, which tend to differ substantially from the intrinsic ones \citep{Giri2018BubbleTomography}. We did, however, not consider the effects of residual foregrounds signals on such measurements. Tests of foreground subtraction techniques do not show any indication of remaining foreground structures being able to mimic late neutral islands. However, such tests are by necessity idealised and the real imaging data will have to used to make a fair assessment of what residual foregrounds will look like.

The population of late neutral islands and the transition into the post-reionization era are a rich source of information about the state of the IGM and the reionization process and deserve more attention. 
We presented a first study based on numerical radiative transfer simulations, but as pointed out above, many aspects deserve further investigation. The future tomographic observations with SKA1-Low, although still some years in the future, make such investigations essential if we want to obtain a complete picture of the EoR.




\section*{Acknowledgements}
We acknowledge Raghunath Ghara and Laura Keating for useful discussions.
This work was supported by the Swedish Research Council (grant number 2016-03581), the Science and Technology Facilities Council (grant numbers ST/F002858/1 and ST/I000976/1) and the Southeast Physics Network (SEPNet). We acknowledge that the results in this paper have been achieved using the PRACE Research Infrastructure resources Curie based at the Trs Grand Centre de Calcul (TGCC) operated by CEA near Paris, France and Marenostrum based in the Barcelona Supercomputing Center, Spain. Time on these resources was awarded by PRACE under PRACE4LOFAR grants 2012061089 and 2014102339 as well as under the Multi-scale Reionization grants 2014102281 and 2015122822. Some of the numerical computations were done on the Apollo cluster at The University of Sussex as well as on resources provided by the Swedish National Infrastructure for Computing (SNIC) at PDC, Royal Institute of Technology, Stockholm. 





\bibliographystyle{mnras}
\bibliography{references,Mendeley}



 \appendix
 \section{Calculation of Euler characteristic using cubical complexes}
In this appendix, we describe how we calculate the Euler chracteristic $\chi$ using cubical complexes.

 \subsection{Construction of cubical complex}
 \label{sec:cubical_complex}
 
 \begin{figure*}
  \centering
  \includegraphics[width=0.9\textwidth]{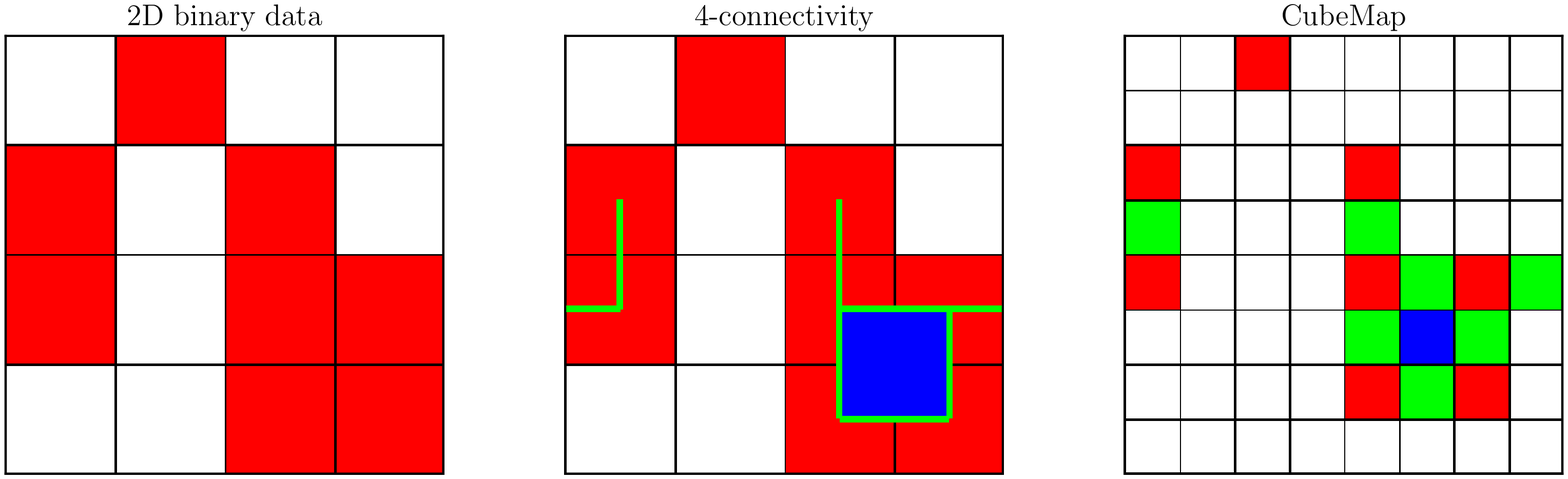}
  \caption{A graphical representation of construction of CubeMap. The origin of the grids shown is at the top left corner. The left most panel shows a 4x4 binary data set where the red cells represent the region of interest and white is the background. In order to construct the cubical complex for this 2D object, we must define a connectivity for the cells. We show the 4-connectivity of the cells in the middle panel. The edges are the lines (green) joining the centres of the cells which are connected based on 4-connectivity. The faces are the surfaces (blue) that can be enclosed by the edges. In the right most panel, we show the CubeMap representation of the object that contains the complete information about the topology. Here, the red, green and blue cells are the points, line segments and squares of the cubical complex respectively.}
  \label{fig:cubemap_cartoon}
\end{figure*}
 
In order to construct a topological measure, such as the $\chi$, we triangulate the structure in our data set using well understood geometrical components. The triangulation method should of course preserve the topological properties of the structure in the data but should also be appropriate for the type of data set. \citet{VanDeWeygaert2011AlphaWeb} worked with the positions of galaxies, so the data set is a cloud of points. The Delaunay tessellation of the cloud of points was found to give a good estimation of the underlying density field \citep{schaap2007delaunay}. Therefore Delaunay triangulation is the proper choice to study the topology of the density field in this case.

Our simulation outputs, on the other hand, consist of 3D data sets defined on equally spaced grids, which is known as digital space. For this case, the appropriate triangulation method is the construction of a so-called cubical complex on our data set. A cubical complex $\boldsymbol{K}$ of any $n$D data set is a set of its $n$-cubes 
which are points, line segments, squares, cubes and so on. 
In the cubical complex, an isolated cell of the data set is represented by a point. Every two connected cells are represented with a line segment. We obtain a square in the complex when any four line segments form a closed structure and a cube when any six squares form a closed structure. To construct a cubical complex, we always need to define the criteria for the grid points or cells to be connected. A more detailed description of cubical complexes can be found in \citet{kaczynski2004computational} and \citet{Wagner2012}.

We use a compact representation of the cubical complex for structures in digital spaces called `CubeMap' \citep{Wagner2012}. It was first introduced in the CAPD library \citep{juda2014capd}. Let us consider an input data of dimension $d$ and size $n^d$, where $n$ is the number of elements in each dimension. The CubeMap representation defined by \citet{Wagner2012} transforms the input data 
to an array with $(2n-1)^d$ elements. 
However, they did not consider their data to be periodic. We extended {CubeMap} to include the periodicity of the data and therefore the array contains $(2n)^d$ elements. CubeMap is a compact way to store topological information of data in a digital space \citep{Wagner2012}. We have implemented this method in our python package to analyse 21-cm signal, {\sc tools21cm}. 



To illustrate how CubeMap works, we consider the example of a binary 2D data set. We show the input $4\times4$ 2D data in the left-most panel of  Fig.~\ref{fig:cubemap_cartoon} where the region of interest is shown in red. We define two cells to be connected when they share a common edge for 2D data (the equivalent in 3D data would be a shared face). 
The connectivity map is shown in the middle panel. The line segments and squares for the cubical complex are shown in green and blue colours respectively. In the right-most panel, we show the corresponding CubeMap array which according to the algorithm has a size $8\times 8$. 

The algorithm can be described as follows. Let $i$ and $j$ be the row and column indices of the cells in any $n\times n$ data set and they vary from 0 to $n$-1. The CubeMap is then constructed using these steps:
\begin{itemize}
 	\item If the cell $(i,j)$ in the data is 1, we fill the cell $(2i,2j)$ in the CubeMap with the value 1 (red in Fig.~\ref{fig:cubemap_cartoon}). These are the points of the cubical complex. For example, the point $(0,1)$ (top row in the example) is 1, which means that the cell $(0,2)$ in the CubeMap becomes 1.
    \item If the cells $(i,j)$ and $(i+1,j)$ in the data are filled, then they are connected along the row according to our definition for connectivity. Therefore we fill the $(2i+1,2j)$ cell in the CubeMap with the value 2, represented by green in Fig.~\ref{fig:cubemap_cartoon}. We perform the same check in the column direction. These cells in the CubeMap array represents the line segments of the cubical complex.
    \item In the CubeMap, if any cell is surrounded by four cells filled with 2s, then that cell represents a square of the cubical complex. We fill these cells in CubeMap with the value 3 (represented by blue).
    \item The last row and column in the CubeMap array only contain line segments and squares when the data set has periodic boundary conditions.
\end{itemize}
The above steps for transforming a 2D data to its CubeMap representation can be easily extended to 3D data. The CubeMap of 3D data will additionally contain cells representing the cubes of the cubical complex, which will be formed if a cell is surround by six squares.

\subsection{Calculating the Euler characteristic from the CubeMap}
 \label{sec:calc_chi}
Once we have constructed a cubical complex from our data set, we can calculate any topological measure. 
Here, we describe the algorithm for calculating the $\chi$ from the CubeMap. $\chi$ is a topological invariant and therefore it does not depend on the cell decomposition or complex used for the structure.

\begin{figure}
  \centering
  \includegraphics[width=0.46\textwidth]{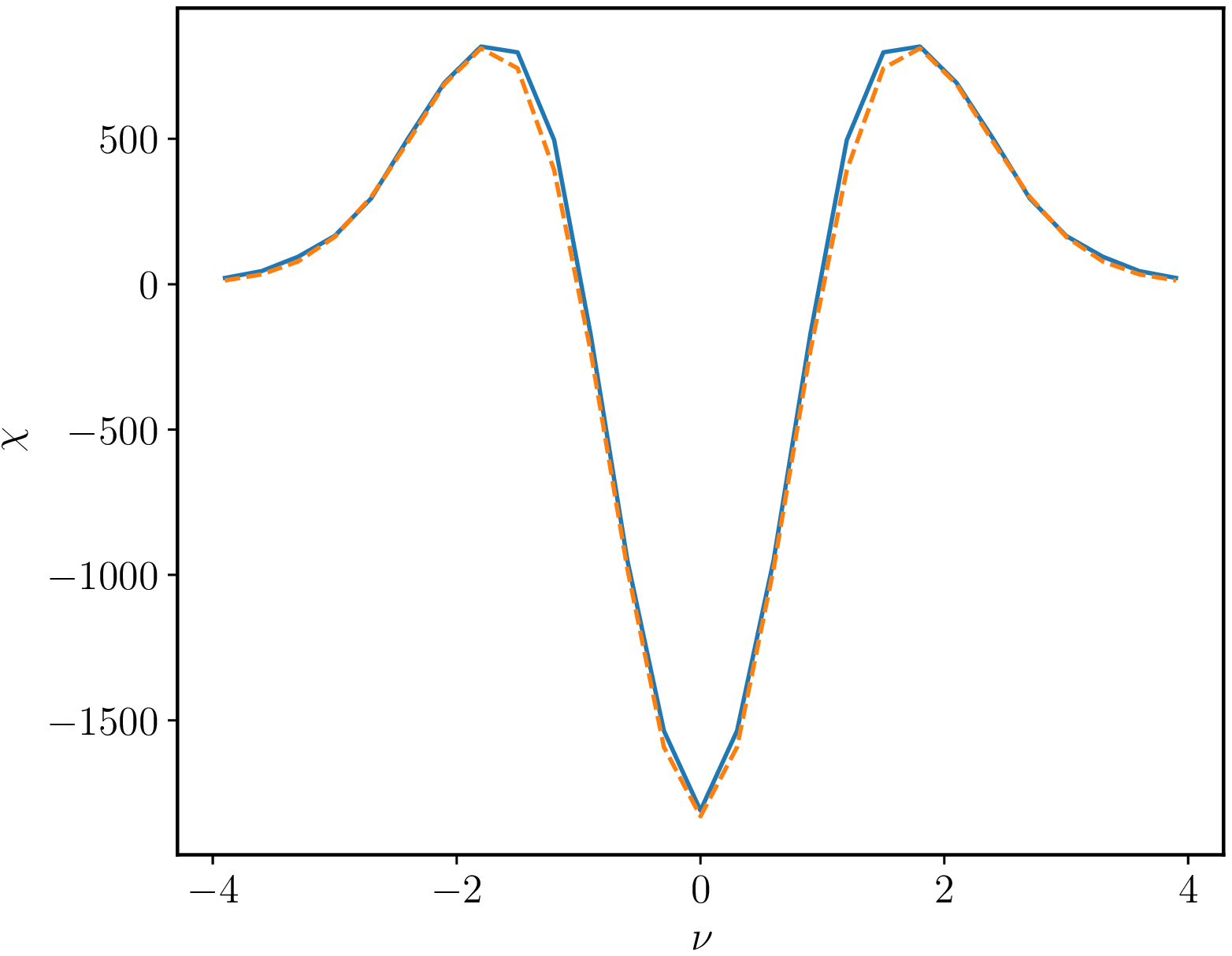}
  \caption{The Euler characteristic ($\chi$) calculated for a Gaussian random field at various threshold $\nu$. The solid and dashed lines show the $\chi$ calculated from our data cube with Gaussian random numbers and equation~(\ref{eq:chi_gauss}) respectively.}
  \label{fig:chi_on_gaussian}
\end{figure}

The $\chi$ of a 3D cubical complex \textbf{K} can be defined as the alternating sum of its $n$-cubes which is as follows,
\begin{eqnarray}
\chi = k_0-k_1+k_2-k_3,
\end{eqnarray}
\noindent where $k_n$ is the number of cubes of dimension $n$. In our CubeMap representation, dimensions of points, line segments, squares and cubes are 0, 1, 2 and 3 respectively. The computation of $\chi$ is thus straightforward after we have converted our data set into CubeMap.

Gaussian fields form a good standard reference to test the sanity of an algorithm for the $\chi$ \citep[e.g.,][]{Schmalzing1997BeyondStructure,Park2013BettiFields}. We create a $(256)^3$ data cube filled with Gaussian random numbers. Due to discretization, our data cube will not be perfectly Gaussian. Therefore, we follow \citet{Park2013BettiFields} and smooth this data with a Gaussian filter with a FWHM of 5 cell widths. In Fig.~\ref{fig:chi_on_gaussian} we plot the value of $\chi$ for different threshold values $\nu$. The symmetric curve shown here is an intrinsic property of a Gaussian field. \citet{doroshkevich1970spatial} and \citet{tomita1990statistics} gave an analytical form for the genus and Minskowski functionals of a Gaussian field. From those we can write the following relation for $\chi$,
\begin{eqnarray}
\chi \propto (\nu^2-1) \exp(-\frac{1}{2}\nu^2) .
\label{eq:chi_gauss}
\end{eqnarray}
We over-plot the analytical curve in Fig.~\ref{fig:chi_on_gaussian} as the dashed line. Our estimated $\chi$ follows the analytical solution very well, thus validating our algorithm.

\section{Toy models}
\label{sec:toy_model}
In this appendix, we explore the relation of the size distributions of ionized regions and neutral regions using a toy model. The size distribution of the ionized regions can be represented by spherical bubbles  embedded in a simulation box. The position and size of the bubbles can be defined by the location and ionizing efficiency of the sources respectively. However the neutral islands are difficult to model with spheres defined by the ionizing sources. 

\begin{figure}
  \centering
  \includegraphics[width=0.46\textwidth]{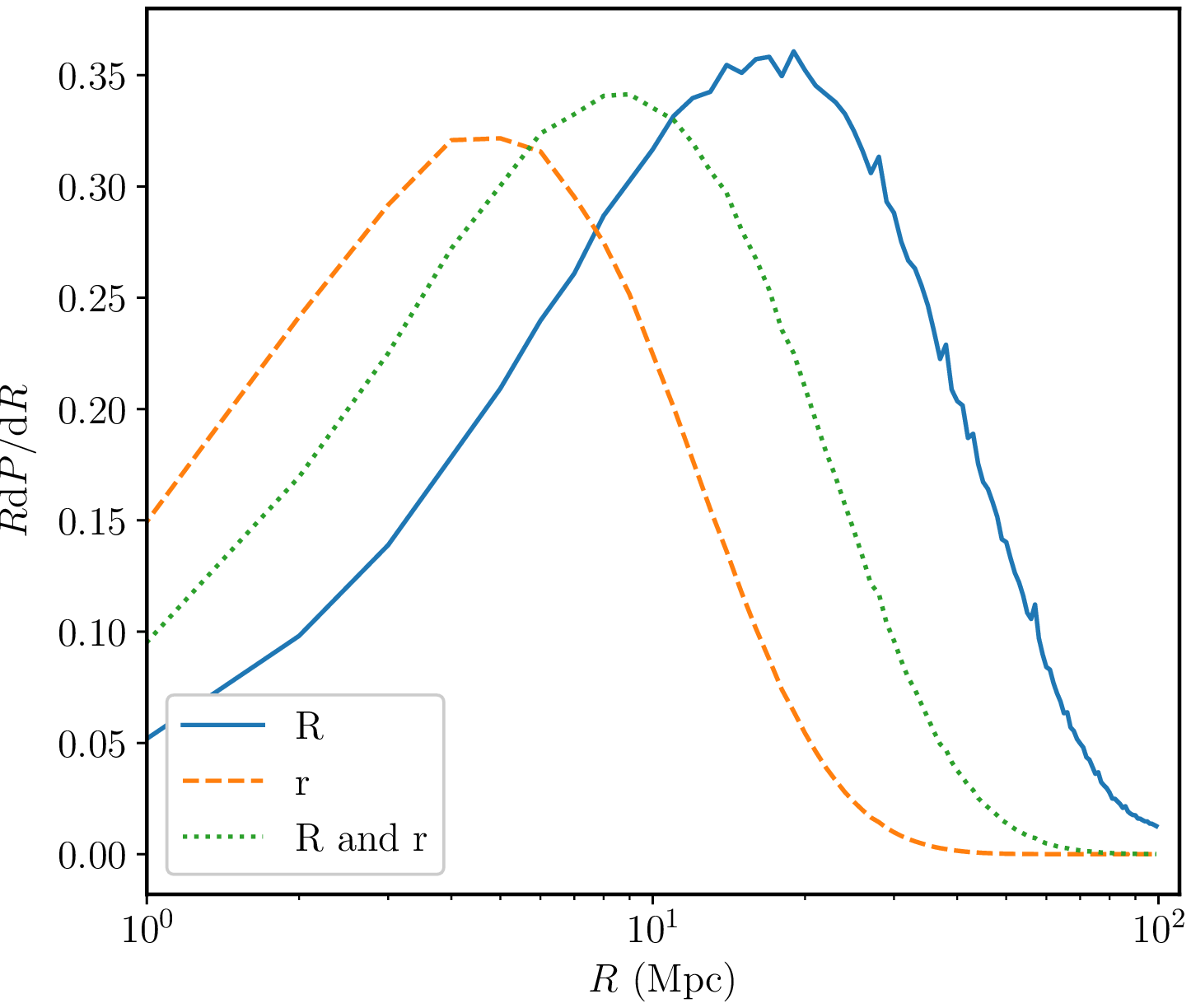}
  \caption{The MFP-ISDs calculated for three different cases from our toy model. All the cases considered here have 70 per cent of the volume filled with ionized regions spheres of radius $R$ = 25 Mpc and $r$ = 5 Mpc.}
  \label{fig:toy_model}
\end{figure}

In order to understand the sizes of the neutral islands, we must visualize the situation in terms of a mathematical optimization problem known as the packing problem \citep[e.g.,][]{Yagiura2017CuttingOptimization.} where we pack ionized bubbled into a simulation box. Let us consider an optimal scenario where the HMACHs form spherical ionized bubbles of identical radii. The Kepler conjecture \citep[e.g.,][]{Hales1998AnConjecture} states that the highest packing fraction achieved is the same as that achieved with hexagonal packing \citep{2003Natur.424...12S}. The size of the islands (or the complementary region of the spheres) depends on the radius of the spheres where smaller spheres will lead smaller voids. When there are smaller spheres along with the larger ones, the smaller spheres can be used to cover the islands created by the larger spheres and thus giving even better packing \citep{1949JAP....20..154H}. This can be proved using the circle packing theorem.

However the ionized bubbles in the Universe will not follow the optimal packing. The regions will overlap and form more complicated configurations. We create a toy model where we put spheres into our simulation box allowing overlapping. We have considered three scenarios. In the first case, we take spheres of radius $R$ = 25 Mpc and keep on placing them randomly into the simulation box until 70 per cent of the volume is filled and then calculate the MFP-ISD of the empty space. We repeat this experiment 100 times to reduce the sample variance and plot the mean MFP-ISD in Fig.~\ref{fig:toy_model}. In the second case, we carry out the same process with smaller spheres with radius $r$ = 5 Mpc. For the last scenario, we considered both kind of spheres. We fill 50 per cent of the box with larger spheres and 20 per cent with smaller ones.

We see that the volume with spheres of larger radius has larger sizes of the neutral islands than that for the box with smaller spheres. The toy model thus validates the inference we made from the hexagonal packing of spheres. We also find that the neutral islands have smaller sizes compared the large spheres only case when we use a mix of smaller and larger spheres. The smaller spheres are more efficient in fragmenting the neutral region of the box.


\bsp	
\label{lastpage}
\end{document}